\documentclass[times]{MyArticle}

\usepackage[colorlinks,bookmarksopen,bookmarksnumbered,citecolor=blue,urlcolor=blue]{hyperref}

\usepackage[usenames,dvipsnames,svgnames,table]{xcolor}
\usepackage{tabularx}
\usepackage{algorithm}
\usepackage{algpseudocode}
\usepackage{graphicx}
\usepackage{graphbox}
\usepackage{subfig}
\newtheorem*{remark}{Remark}

\algrenewcommand\algorithmicrequire{\textbf{Input:}}
\algrenewcommand\algorithmicensure{\textbf{Output:}}

\begin{document}

\title{Adaptive coupling of 3D and 2D fluid flow models}

\author{Pratik Suchde}

\address{University of Luxembourg, Luxembourg \break Fraunhofer ITWM, Germany%
}
\corraddr{E-mail: pratik.suchde@gmail.com, pratik.suchde@uni.lu}

\begin{abstract}
Similar to the notion of h-adaptivity, where the discretization resolution is adaptively changed, I propose the notion of model adaptivity, where the underlying model (the governing equations) is adaptively changed in space and time. Specifically, this work introduces a hybrid and adaptive coupling of a 3D bulk fluid flow model with a 2D thin film flow model. As a result, this work extends the applicability of existing thin film flow models to complex scenarios where, for example, bulk flow develops into thin films after striking a surface. At each location in space and time, the proposed framework automatically decides whether a 3D model or a 2D model must be applied. Using a meshless approach for both 3D and 2D models, at each particle, the decision to apply a 2D or 3D model is based on the user-prescribed resolution and a local principal component analysis. When a particle needs to be changed from a 3D model to 2D, or vice versa, the discretization is changed, and all relevant data mapping is done on-the-fly. Appropriate two-way coupling conditions and mass conservation considerations between the 3D and 2D models are also developed. Numerical results show that this model adaptive framework shows higher flexibility and compares well against finely resolved 3D simulations. In an actual application scenario, a 3 factor speed up is obtained, while maintaining the accuracy of the solution. 
\end{abstract}



\maketitle

\vspace{-6pt}


\section{Introduction}
\label{sec:Intro}

Flow phenomena that combines fluid dynamics on a surface and that in the bulk occur in many fields, ranging from spray coating to machine lubrication and tyre aquaplaning. To simulate such phenomena, various specialized thin film flow models have been developed that accurately capture surface-level flow effects. Popular thin film flow models include the shallow water equations \cite{vreugdenhil2013numerical} and the lubrication approximation \cite{bertozzi1996lubrication}, among many others \cite{chakraborty2014extreme, fries2018higher, martin2023physics, rohlfs2018wavemaker}. All film flow models reduce the dimensionality of the governing equations, making them more efficient than bulk flow models for simulating fluids in a thin layer. 

Such film models are typically applied to situations where a fluid film is the only fluid of interest, or where a fluid film forms an immiscible interface between two bulk fluids. However, these models have not been applied when there is a dynamic interplay between bulk and film flow. In a scenario where bulk flow develops into a thin film, the thin layer of fluid is usually treated just as any other bulk three-dimensional fluid region, using full 3D models. Simulations of such scenarios typically rely on significantly finer resolutions in the regions with thin film flow, which dramatically increases computation time. This makes full dynamic simulations unrealistic for actual applications. As a result, advances made in modelling thin fluid films have not been fully utilized in complex applications involving both surface and bulk flow. 

In the present work, I propose a simulation framework for scenarios where bulk (3D) and thin film flow (2D) occur beside each other. The proposed algorithms can choose on-the-fly when and where a thin film flow model is applicable, and can shift the governing equations automatically between a 3D flow model and a 2D model. Consider the example of a cleaning process in the food industry, where a large tray with food remnants is being cleaned with a jet of water (see Section~\ref{sec:Cleaning} for details). The high velocity and flow rates of the jet would typically require a 3D flow model, like the Navier--Stokes equations, to be modelled accurately. As the jet hits the solid surface being cleaned, it would form a thin layer of fluid, which could be simulated quickly with a 2D thin film flow model. If sufficient fluid collects on the surface, the assumptions of a thin film model would no longer be applicable, and a 3D bulk flow model would be required again. Realizing this idea forms the main goal of the present work.


Coupling bulk and thin film flow has been widely done for modelling a tsunami landing on a shoreline (for example, \cite{ling2023two,mintgen2018bi, pan2024variable, pan2021mpm}). Here, a thin film model, typically the shallow water equations, are used to model the flow of the ocean sufficiently far away from the coast, under the assumption that the ocean is significantly wider than deep in these regions. Near the coastline, the tsunami wave is modelled with a bulk flow model, typically the Navier--Stokes equations. The present work generalizes this existing literature by removing several key underlying assumptions present in tsunami modelling: 
\begin{enumerate}
    \item There is a fixed interface between the regions of the domain where the thin film model and the bulk flow model are applied.
    \item This fixed interface can be prescribed a-priori, before starting the simulation. 
    \item The thin film flow is fully developed. 
\end{enumerate}
The present work removes these three assumptions by coupling 2D and 3D models without prescribing a fixed interface between them. The interface(s) between the two models will be automatically determined by newly developed algorithms, such that the location and presence of an interface can change in both space and time. This will allow moving interfaces, and even allow scenarios where the interface may completely disappear at an instant of time. This implies that it is possible for the entire simulation domain to be modelled by only a thin film model, or only a bulk flow model at a particular time step. This framework is referred to as being model adaptive, since the model (governing equations) being used are adaptively changed during the simulation. 

The two models coupled in this work are (i) 3D bulk flow governed by the incompressible Navier--Stokes equations, discretized with an implicit meshfree collocation approach \cite{suchde2018meshfree}, and (ii) a pseudo-2D particle based thin film flow model called the discrete droplet method \cite{bharadwaj2022discrete}, with an explicit Lagrangian solution scheme that models film flow by tracking the movement of droplets on a surface. The framework introduced here can be easily adapted to be used with any Lagrangian meshfree or particle-based fluid solver and thin film solver. The key novelty and emphasis here is on the notion of model adaptivity, and not the choice of models or discretization methods.

\begin{remark}
    The thin film flow model is referred to as \emph{pseudo}-2D since it also covers flow over curved surfaces where a locally 2D model is used. For the sake of brevity, I drop the term pseudo, and refer to the thin film model as a 2D model. 
\end{remark}

It is important to note here that not only are the governing equations different in the two models, but the discretization itself is different as well. In the bulk flow approach, the entire flow domain is discretized. On the other hand, in the thin film model, only the surface is discretized, with the depth of the fluid film determined using an additional governing equation. Thus, in addition to being \emph{model adaptive}, this framework is also \emph{discretization adaptive}. 




The paper is organized as follows. To start the paper, Section~\ref{sec:Preliminaries} introduces some necessary preliminaries without any novel work. This includes the terminology, and a brief summary of the bulk and thin film flow models used in the present work. Then, Section~\ref{sec:Overview} presents an overview of the novel model adaptivity framework. The framework is split into three parts, which form the next three sections: Section~\ref{sec:Where} proposes how to determine where the underlying model needs to be changed, Section~\ref{sec:Transition} presents how to change the model being used
, and Section~\ref{sec:DataComm} presents how data can be communicated between two models that occur beside each other. Subsequently, numerical results, validations and applications are shown in Section~\ref{sec:Results} followed by a discussion on the limitations of the work in Section~\ref{sec:Limitations}. The paper is then concluded with an outlook and summary in Section~\ref{sec:Conclusion}.


\section{Preliminaries}
\label{sec:Preliminaries}

In this section, I introduce both the bulk flow model, and the thin film model, and related preliminaries. I once again emphasize that the novelty in this work is not in the models themselves, but is in how the models are combined in an adaptive manner. 

Throughout this work, I only consider domains in $\mathbb{R}^3$, with a three-dimensional bulk flow model, and a two-dimensional thin film model. However, for the ease of visualization, several schematics and figures shown in the methodology sections \ref{sec:Overview}--\ref{sec:DataComm} illustrate domains in $\mathbb{R}^2$, with a two-dimensional bulk flow model and a one dimensional thin film model.

\begin{remark}
    Since the 3D and 2D models will be solved one after the other, the model adaptivity and coupling presented in this work is independent of the spatial and temporal discretization of the individual models. 
\end{remark}


\subsection{Nomenclature}

For the sake of clarity, this subsection defines the key terminology used throughout this work. 
\begin{itemize}
    \item Throughout this work, the term \textit{model} refers to the governing equations, with relevant initial and boundary conditions. Thus, the two models being used are a 3D bulk flow model, and a 2D thin film flow model. 

    \item \textit{Model adaptivity} refers to the process of adaptively choosing the model during a simulation. Note that this means a complete change of the governing PDEs, not just dropping one term from a PDE, as the term has been used in literature (for example, \cite{behrens2019towards, braack2003posteriori}).


    \item \textit{Model transition} refers to the process of change or transition from one model to another. So either from the 3D model to the 2D one, or vice versa. 
\end{itemize}


\subsection{Meshfree terminology}
\label{sec:Meshfree}

The use of two meshfree methods, as opposed to mesh-based approaches, is crucial to the present work as it is straightforward to switch the underlying model being solved at a particle. Furthermore, meshfree methods also provide an easy approach for simulating free surface flow, which is central to the 2D-3D simulations considered here. 

Consider a computational domain $\Omega = \Omega(t)$ partitioned into two non-overlapping regions, 
$\Omega(t) = \Omega^{\text{3D}}(t) \cup \Omega^{\text{2D}}(t)$ 
where $\Omega^{\text{3D}}$ is the subdomain where the 3D bulk flow model is being applied, 
and $\Omega^{\text{2D}}$ is the subdomain where the 2D thin film flow model is being applied. Both $\Omega^{\text{3D}}$ and $\Omega^{\text{2D}}$ could contain free boundaries. The computational domain is discretized with a cloud of $N^{\text{total}} = N^{\text{2D}} + N^{\text{3D}}$ points or particles, with $N^{\text{2D}} = N^{\text{2D}}(t)$ discretizing $\Omega^{\text{2D}}$, and $N^{\text{3D}} = N^{\text{3D}}(t)$ discretizing $\Omega^{\text{3D}}$. The terms point and particle are equivalent for the present work. 

\begin{remark}
If the entire domain is governed with a bulk flow model at a particular instant of time $t^*$, we would have $N^{\text{2D}}(t^*)=0$. Similarly, $N^{\text{3D}}(t^*)=0$ holds if only the thin film flow model is being used at that time. 
\end{remark}

For both 3D and 2D domains, for a particle $i$, all approximations will be carried out on a compact support $S_i$ of nearby particles, also referred to as its neighbourhood 
\begin{equation}
\label{Eq:Nbd}
    S_i = \left\{ \vec{x}_j | \, \| \vec{x}_j - \vec{x}_i \| \leq \frac{h_i + h_j}{2} \right\} \,,
\end{equation}
where $h_i = h(\vec{x}_i)$ is the support radius or interaction radius at particle $i$. Note that $i \in S_i$. For more details on the support definitions, and other related meshfree basics, I refer the reader to \cite{suchde2017flux} for 3D domains, and \cite{bharadwaj2022discrete, suchde2019fully} for 2D domains and curved surfaces. The point cloud discretizing the initial domain is generated using a meshfree advancing front technique \cite{lohner1998advancing, slak2019generation, suchde2023volume}.


\subsection{Bulk flow model}
\label{sec:3Dmodel}

The bulk flow model used is the three-dimensional incompressible Navier--Stokes equations in a Lagrangian framework. 
\begin{align}
    \nabla \cdot \vec{v} &= 0 \,, \\
    \frac{D \vec{v}}{Dt} &= \frac{\eta}{\rho} \Delta \vec{v} - \frac{1}{\rho} \nabla p + \Vec{g} \,,
\end{align}
where $\vec{v}$ is the flow velocity, $p$ is the pressure, $\rho$ is the density, $\eta$ is the dynamic viscosity, and $\frac{D}{Dt}$ is the material derivative. For the sake of simplicity, temperature and turbulent effects are not considered. Throughout this work, this bulk flow model is also referred to as the 3D model for short. 

\subsubsection*{Temporal discretization}~\\
Time discretization is done using a modified projection scheme \cite{chorin1968numerical}. The scheme starts with Lagrangian motion of particles \cite{suchde2018point}. This is followed by point cloud organization to maintain a quasi-regular point cloud and to prevent distortion \cite{suchde2023point, suchde2019fully}. Subsequently, an implicit intermediate velocity is determined, which is then projected to a divergence free space using an implicit pressure Poisson equation followed by the pressure update. For further details on the numerical scheme, I refer to~\cite{michel2021meshfree, suchde2017flux, suchde2018meshfree}.

\subsubsection*{Spatial discretization}~\\
The 3D domain is discretized using a meshfree method, as described in Section~\ref{sec:Meshfree}. The particles discretizing the domain can be of three types: (i) wall particles that lie on a solid wall, (ii) free surface particles, and (iii) interior particles that do not belong to any boundary. 

The inter-particle distances are intrinsically linked to the interaction radius, as done in \cite{Drumm2008, SeiboldThesis, suchde2017flux}. The minimum and maximum distance between two nearby particles, also referred to as separation and fill distance respectively, is fixed at $r_{\text{min}} h$ and $r_{\text{max}} h$ for interaction radius $h$ and fixed parameters $r_{\text{min}}$ and $r_{\text{max}}$. In a moving Lagrangian framework, these distance criteria are enforced through the addition and deletion of particles. I refer to \cite{suchde2023point, suchde2019fully} for more details. Thus, $h$ serves as both the interaction radius and the resolution of the discretization. 

Derivatives are approximated using a meshfree collocation approach, referred to as the Generalized Finite Difference Method (GFDM) \cite{benito2001influence, rao2023upwind, zheng2022theoretical}. Unlike several other meshfree methods, the GFDM ensures discrete consistency up to the desired order of accuracy. This means that monomials up to the prescribed order of accuracy are differentiated exactly, even at the discrete level. For more details on GFDMs, I refer the reader to \cite{jacquemin2020taylor, suchde2019meshfree}. Throughout this work, an order of accuracy of $2$ is prescribed. 



\subsection{Thin film flow model}
\label{sec:2Dmodel}

There are two important considerations in the choice of the thin film flow model
\begin{enumerate}
    \item The model should be able to capture the formation of thin fluid films, not just pre-existing films. This is essential for applications that consider bulk flow hitting a surface and forming a thin fluid film; see, for example, Section~\ref{sec:Cleaning}. 
    \item The model should allow free boundaries within the thin film flow. 
\end{enumerate}
Based on this, I choose the recently proposed discrete droplet method (DDM) \cite{bharadwaj2022discrete, bharadwaj2024lagrangian} as the thin film flow model. The DDM is a meshfree approach that models incompressible flow in a thin fluid film using a Lagrangian framework by tracking the movement of individual fluid droplets. The momentum conservation equation is given by
\begin{equation}
\label{Eq:FilmMomentum}
    \frac{D\vec{v}}{Dt} = -\frac{\eta}{\rho H_{\text{film}}^2} \vec{v} - \frac{1}{\rho} \nabla p + \vec{g}_{\|}  \,,
\end{equation}
%
%
where $\vec{v}$ is the velocity parallel to the surface on which the fluid is flowing, $p$ is the pressure, $\vec{g}_{\|}$ is the component of the gravity parallel to the surface, $\rho$ is the density, $\eta$ is the dynamic viscosity, and $H_{\text{film}}$ is the height or depth of the fluid film. The terms height and depth are used interchangeably in this work. The height of the fluid film is computed based on the aggregation of droplets. The height of the film at a particle $i$ is given by 
\begin{equation}
\label{Eq:Height}
    H_{\text{film}, i} = \sum_{j \in S_i}  \frac{\alpha}{\pi h_j^2} \exp\left(-\alpha\frac{\|\vec{x_j}-\vec{x_i}\|^2}{h_j^2}\right) V_j \,,
\end{equation}
where $h_j$ is the support radius of particle $j$, $S_i$ is the support or neighbourhood of particle $i$ (see Section~\ref{sec:Meshfree}), and $V_j$ is the volume of particle $j$. Each particle is assumed to a droplet of diameter $d$. Thus, we have $V_j = \frac{1}{6} \pi d_j^3$. The coefficient $\alpha$ is computed from a mass conservation condition, see~\cite{bharadwaj2022discrete}. The mass of a droplet particle $i$ is given by $m_i = \rho_i V_i$. Note that both mass conservation and the divergence-free velocity condition are implicit to the model described here, see \cite{bharadwaj2022discrete}. Throughout this work, this thin film flow model is also referred to as the 2D model for short. 

\subsubsection*{Temporal discretization}~\\
An explicit temporal discretization is carried out for the thin film model. Time integration starts with the Lagrangian motion of droplets, followed by an explicit update of the momentum equation (Eq.\,\eqref{Eq:FilmMomentum}). Subsequently, the updated height function (Eq.\,\eqref{Eq:Height}) is determined at the new droplet locations. For more details, I refer the reader to \cite{bharadwaj2022discrete}.

\subsubsection*{Spatial discretization}~\\
The domain discretization of the regions where the thin film model is being applied is done using particle or droplets, as explained above. A key point to note here is that only the surface is discretized, not the bulk. Droplets move on the surface, with the height of the film built up as described in Eq.\,\eqref{Eq:Height}.

Spatial derivatives are computed using a GFDM approach, just as the bulk flow model. I refer to~\cite{suchde2019fully, suchde2019meshfree}, for details on computation of GFDM spatial derivatives on surfaces.



\section{Model adaptivity overview}
\label{sec:Overview}

For the initial discretization and model selection, I assume that the user prescribes where in the domain the 3D and 2D models will be applied. Note that the user could also prescribe only one of the two models being present as the initial condition. Now consider the simulation time $t = t^*$, when either one or both models are present in the domain. The novel procedure to adaptively change the underlying model during a simulation is split into three questions 
\begin{enumerate}
    \item \emph{Where} in the domain does the model need to be changed? This is answered in detail in Section~\ref{sec:Where}.
    \item In the parts of the domain where a desired model change is identified, \emph{how} should the model transition be carried out? This question is tackled in Section~\ref{sec:Transition}.
    \item Where both models occur beside each other, how to ensure \emph{data communication} between the models? This issue is addressed in Section~\ref{sec:DataComm}.
\end{enumerate}
\begin{remark}
    In existing literature coupling 3D and 2D flow that uses a fixed and pre-defined interface between the models, like that in tsunami modelling literature \cite{ling2023two,marras2020modeling, maso2022lagrangian}, only the third question is relevant. The first two questions are specific to the model adaptivity framework introduced here. Furthermore, for tsunami modelling, the solution to the third question relies on a pre-processing step before the simulation is started. In contrast, here, the data communication needs to be dynamic without any pre-processing possible, since the interface between models is not fixed.
\end{remark}

\begin{figure}
  \centering
  \subfloat[Initial condition: The user prescribes where the 3D bulk model should used (blue), and where the 2D thin film flow model should be used (maroon). ]{\includegraphics[align=c,width=0.8\textwidth]{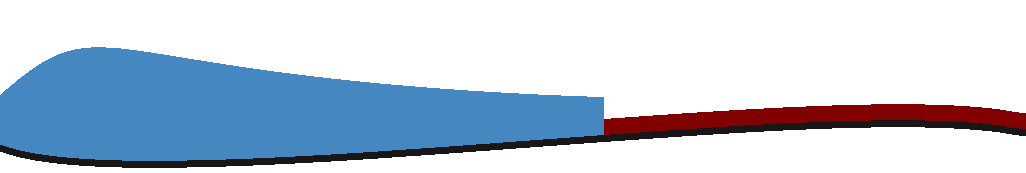}} \\  
  \subfloat[Model adaptivity step 1: Identify where the model needs to be changed (regions marked in green). See Section~\ref{sec:Where} for details.]{\includegraphics[align=c,width=0.8\textwidth]{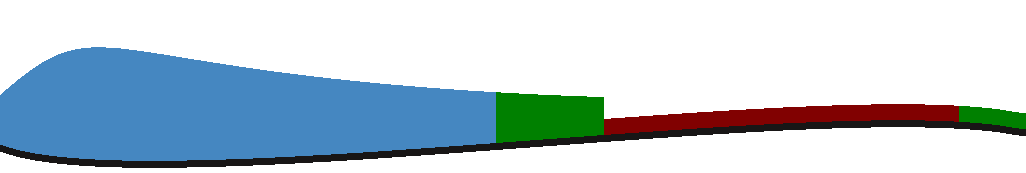}} \\
  \subfloat[Model adaptivity step 2: Transition from one model to the other (based on regions identified in Step~1). See Section~\ref{sec:Transition} for details.]{\includegraphics[align=c,width=0.8\textwidth]{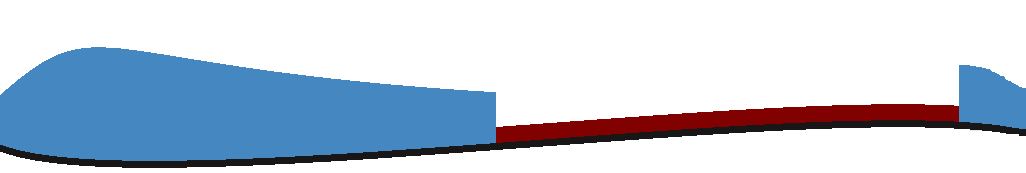}} \\ 
  \subfloat[Model adaptivity step 3: Create handshake or buffer region for data communication between the two models (see Section~\ref{sec:DataComm} for details). The domain is shown with a high transparency. The buffer regions are are shown in a brighter shade of the respective models: blue indicates the buffer region for the bulk models, and red indicates the buffer region for the thin film model.]{\includegraphics[align=c,width=0.8\textwidth]{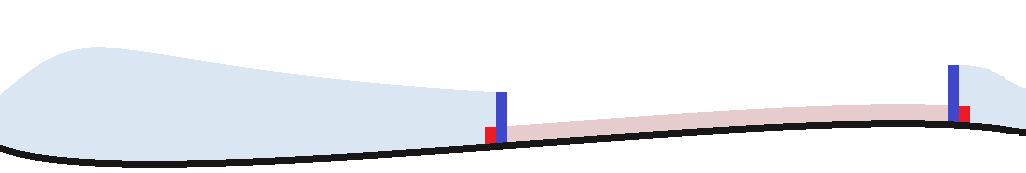}} \\   
  \caption{Schematic of model adaptivity between a 3D bulk flow model (blue)  and 2D thin film flow model (maroon). The surface over which the fluid is flowing is shown in black. The thin film flow model is shown to be of a smaller height since it relies on a surface discretization only, while the entire bulk is discretized for the bulk flow model. Throughout this work, the bulk flow model is three-dimensional, while the thin film model is two-dimensional. Lower dimensional representations are used in this figure for ease of visualization.}
  \label{Fig:Overview}%
\end{figure}

Before each of these three questions are addressed individually in the coming three sections, I first present an overview of the numerical scheme in a schematic in Figure \ref{Fig:Overview} and in Algorithm~\ref{alg:Overview}. 

\begin{algorithm}
    \caption{Model adaptivity overview: Numerical solution scheme} \label{alg:Overview}
    \begin{algorithmic}
        \State User-defined initial domain discretization
        \While{Time-stepping loop}
            \State Check where the model needs to be changed \Comment{See Section~\ref{sec:Where}}
            \If{Model transition detected}
                \State Perform model transition \Comment{See Section~\ref{sec:Transition}}
                \State Delete buffer region from last time step \Comment{See Section~\ref{sec:DataComm}}
                \State Create new buffer region \Comment{See Section~\ref{sec:DataComm}}
            \EndIf
            \State Point cloud organization: addition, deletion of particles
            \State Bulk flow solver \Comment{See Section~\ref{sec:3Dmodel}}
            \State Thin film flow solver \Comment{See Section~\ref{sec:2Dmodel}}
            \State Post-processing calculations
        \EndWhile
    \end{algorithmic}
\end{algorithm}


%
%


\section{Detecting model transition}
\label{sec:Where}

This section introduces the automatic domain decomposition into the subdomains $\Omega^{\text{3D}}$ and $\Omega^{\text{2D}}$ for the 3D and 2D models respectively. The initial definition of subdomains is user-prescribed. Subsequently, rather than performing a full domain decomposition at each time step, I only check for regions in the domain where the model being used needs to be changed from 2D to 3D, or vice versa. This event of changing the model is referred to as \emph{model transition} throughout the present work. Before introducing the model transition criteria, I first establish a set of conditions that need to be fulfilled:
\begin{enumerate}
    \item \label{item:Local} \emph{Locality:} Any criterion to detect model transition must be local at each particle. For a particle $i$, the detection of model transition must be made based only on its neighbourhood $S_i$. This will avoid global checks, and make the detection more efficient. 
    \item \label{item:Resolution} \emph{Resolution:} The choice of model should take into account the user-prescribed minimum (finest) resolution of the 3D model. 
    \item \label{item:Flexible} \emph{Flexible:} Additional user-defined criteria for model selection should be allowed. 
\end{enumerate}
All particles are evaluated separately to determine if model transition is needed. Particles that satisfy all criteria presented below are flagged for transition. The actual transition of particles from one model to another, see Section~\ref{sec:Transition}, is only done after all particles are evaluated. I further split the detection of model transition into two parts:
\begin{enumerate}
    \item Detecting when a 3D particle needs to be transitioned to a 2D particle, see Section~\ref{sec:BulkToFilm}.
    \item Detecting when a 2D particle needs to be transitioned to a 3D particle, see Section~\ref{sec:FilmToBulk}.
\end{enumerate}
%


\subsection{Detecting 3D to 2D transition}
\label{sec:BulkToFilm}



Consider a 3D boundary particle $i$ on a solid wall. Since the 2D thin film approximation is only applicable on and near walls, only wall particles need to be checked for transition to a 2D model. I now present a series of criteria, each of which need to be satisfied for a 3D particle to be flagged for model transition.

\subsubsection{Resolution}
\label{sec:Resolution_3Dto2D}
\begin{figure}
  \centering
  \subfloat[The particle being evaluated has only interior and wall boundary neighbours, with no free surface neighbours. No transition to thin film will occur.]{%
  \includegraphics[align=c,width=0.48\textwidth]{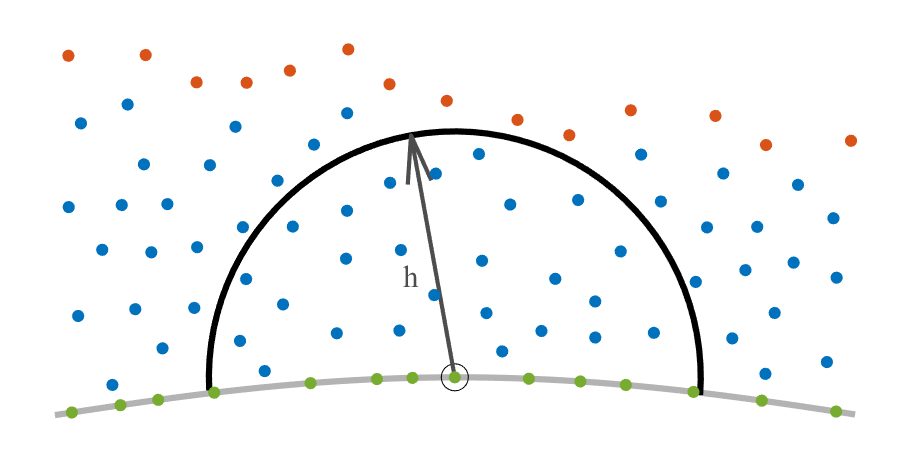}
  \label{Fig:Resolution_a}} \hspace{0.02\textwidth}
  \subfloat[The particle being evaluated has free surface neighbours (in addition to interior and wall boundary neighbours). The PCA criterion will be evaluated to determine if transition to thin film will occur.]{%
  \includegraphics[align=c,width=0.48\textwidth]{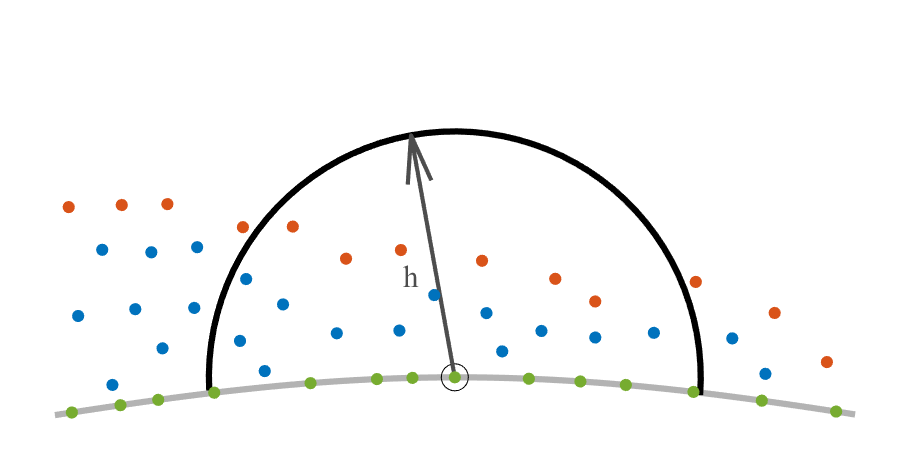}
  \label{Fig:Resolution_b}}\\
  \subfloat[The particle being evaluated has no interior or free surface neighbours. This particle will be changed to the thin film model without the need to evaluate the PCA criterion.]{%
  \includegraphics[align=c,trim={0 0 0 2cm}, clip, width=0.48\textwidth]{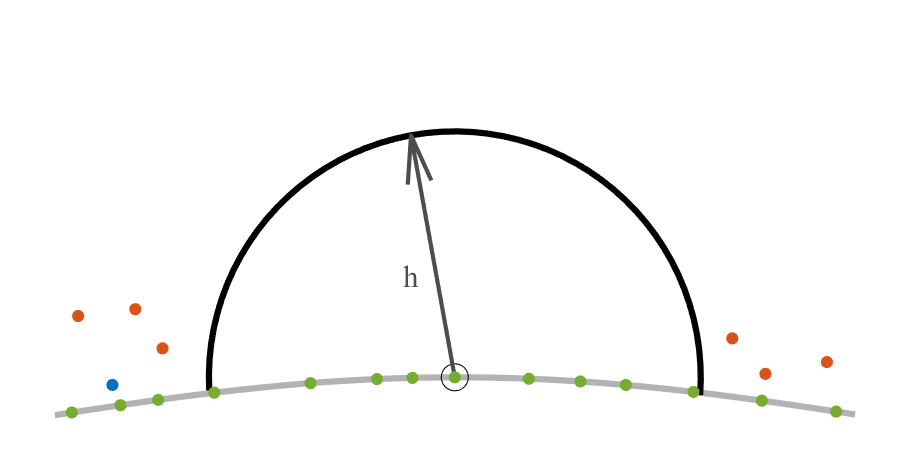}
  \label{Fig:Resolution_c}}
	\caption{Resolution based criterion for detecting bulk to thin film transition. Wall particles are marked in green, interior particles in blue, and free surface particles in orange. The wall particle at which the bulk to thin film transition criterion is being evaluated is marked with an additional black circle, with its neighbourhood highlighted. The wall is marked in grey. Note that throughout this work a 3D bulk model and 2D thin film model is considered. This figure shows a lower dimensional schematic for the ease of visualization.}
  \label{Fig:ReslutionCriterion}%
\end{figure}

The first criterion to be considered is based on condition \ref{item:Resolution} of resolution dependence. I assume a user-defined resolution for the 3D model, which could vary in both space and time. If the resolution at particle $i$ is deemed insufficient to apply the 3D model, then the particle is flagged for transition to the 2D model. Conversely, if the resolution is deemed sufficient for the 3D model, then model transition is not done at this particle, and further criteria of model transition (see below) are not evaluated. This creates a condition between the resolution of the 3D particle and the height of the thin film model if it were to be applied at that location. This notion of checking the resolution is evaluated by looking for free surface particles close to a wall. 
\begin{itemize}
    \item If the 3D particle $i$ has interior neighbours, but no free surface neighbours, then no model transition will be done. All neighbours of such a particle would either be interior particles or wall boundary particles, see Figure~\ref{Fig:Resolution_a}, and particles are present throughout the neighbourhood. Thus, the effective height of a theoretical thin film model at this location would be larger than the interaction radius: $h_i < H_\text{film}(\vec{x}_i)$, making it impossible to prescribe a local condition for model transition. 
    \item If particle $i$ has at least one free surface neighbour, then it is flagged for further checks to see if a thin film model is applicable, see Figure~\ref{Fig:Resolution_b}. This implies that the bulk model resolution may not be sufficiently fine to capture the fluid behaviour at that location: $h_i > H_\text{film}(\vec{x}_i)$.
    \item If particle $i$ only has wall boundary neighbours, with no interior or free surface neighbours, it is flagged for model transition to the thin film model, see Figure~\ref{Fig:Resolution_c}. We have $h_i \gg H_\text{film}(\vec{x}_i)$.
\end{itemize}

As explained in Section~\ref{sec:3Dmodel}, the minimum and maximum inter-particle distance are a fixed ratio of the interaction radius. Thus, $h$ is also an indicator of the resolution of the point cloud. As a result, the criterion of looking for free surface and interior particle neighbours within the interaction radius is inherently based on the resolution. 
    
\begin{remark}
    In contrast to the meshfree collocation approach used here, in several meshfree methods like Smoothed Particle Hydrodynamics (SPH), there is usually no fixed relation between the interaction radius and the minimum particle spacing. Thus, this resolution criteria would need to be modified if this framework is to be applied to an SPH based bulk discretization. 
\end{remark}







\subsubsection{PCA: eingenvalues}
\label{sec:PCA1}

The fundamental assumption in all thin film models is that the fluid has a depth much smaller than the characteristic length scale in the other directions. This assumption is key in determining whether a thin film model is applicable at a particular location. Thus, the applicability of a thin film model could be investigated by looking at the local geometric structure of the point cloud. 

I investigate the applicability of a thin film model through a local application of Principal Component Analysis (PCA), which has been widely used for linear structure approximation of scattered data. To perform local PCA on a particle $i$ with neighbourhood $S_i$, consider the eigenvalues of the local covariance matrix defined as
\begin{equation}
	\mathbf{P}_i = \sum_{j \in S_i} 
		  \left( \vec{x}_j - \vec{x}_c \right) 
		  \left( \vec{x}_j - \vec{x}_c \right)^T \,,
\end{equation}
where $\vec{x}_j$ is a column vector prescribing the location of particle $j$, and $\vec{x}_c$ is the centroid of all the neighbouring particles
\begin{equation}
	\vec{x}_c = \frac{1}{n(S_i)} \sum_{ j \in S_i } \vec{x}_j \,,
\end{equation}
with $n(S_i)$ being the number of particles in the neighbourhood $S_i$. I refer the reader to \cite{Shlens2014} for more details on PCA computation. 

\begin{remark}
    PCA has been successfully applied to prescribe a local coordinate system for point cloud surfaces and for surface normal computation (for example, \cite{Hoppe1992, Liang2013}). For point cloud surfaces or manifolds, the relative sizes of the eigenvalues of the covariance matrix $\mathbf{P}_i$ reveals information about the (local) dimensionality of the manifold. For a surface in $\mathbb{R}^3$, one eigenvalue will necessarily be much smaller than the other two
\begin{equation}
\label{Eq:Eigen}
    \lambda_1 > \lambda_2 \gg \lambda_3
\end{equation}
The eigenvectors corresponding to $\lambda_1$ and $\lambda_2$ will span the tangent space, while the eigenvector corresponding to $\lambda_3$ will be a surface normal vector. 
\end{remark}

Now, I extend this work on PCA for point cloud surfaces \cite{Hoppe1992, Liang2013} to determine the applicability of a thin film model. This relies on the assumption that a bulk point cloud discretization of a thin film has similar structural properties to a point cloud discretizing a surface. While Eq.\,\eqref{Eq:Eigen} may not hold, the relative sizes of the eigenvalues can be used to devise a measure to examine the local point cloud. Consider the eigenvalues $\lambda_1 > \lambda_2 > \lambda_3$. Heuristically, I propose a criterion that if $\lambda_3$ is ``small enough", the model for particle $i$ should be transitioned to a thin film approximation. Quantitatively, we need to define a threshold in the differences between $\lambda_1$, $\lambda_2$ and $\lambda_3$. For this, I consider the ratio
\begin{equation}
    \lambda_3 = \epsilon_{\lambda} \frac{\lambda_1 + \lambda_2}{2}\,.
\end{equation}
For a point cloud discretizing a 2-manifold, we have $\epsilon_\lambda \ll 1$. The criteria of $\lambda_3$ being ``small enough" for the applicability of a thin film model is realized by imposing a maximum value of $\epsilon_\lambda$. If 
\begin{equation}
\label{Eq:EigenvalueLimit}
    \epsilon_\lambda < \epsilon_\lambda^{\text{max}} \,,
\end{equation}
I flag particle $i$ for model transition. The value of $\epsilon_\lambda^{\text{max}}$ is set by numerical trial and error to $0.15$.

\subsubsection{PCA: eigenvectors}
\label{sec:PCA2}

\begin{figure}
  \centering
  \subfloat[None of the eigenvalues is significantly larger. No transition to the film model will take place.]{%
  \includegraphics[align=c,width=0.48\textwidth]{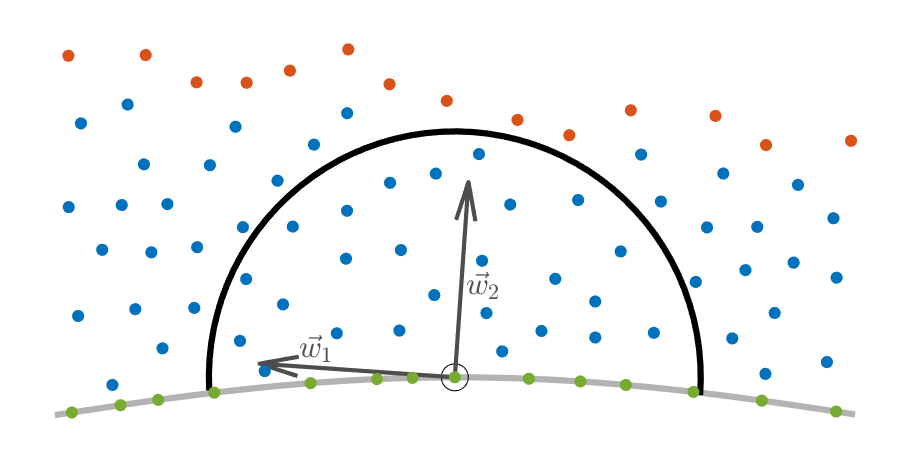}
  \label{Fig:EigenVector_a}} \hspace{0.02\textwidth}
  \subfloat[One eigenvalue (corresponding to eigenvector $\vec{w}_1$) is much larger than the other eigenvalue (corresponding to eigenvector $\vec{w}_2$). Furthermore, the eigendirection $\vec{w}_2$ is ``close" to the normal $\vec{n}$. This particle will thus be transitioned to the thin film model.]{%
  \includegraphics[align=c,width=0.48\textwidth]{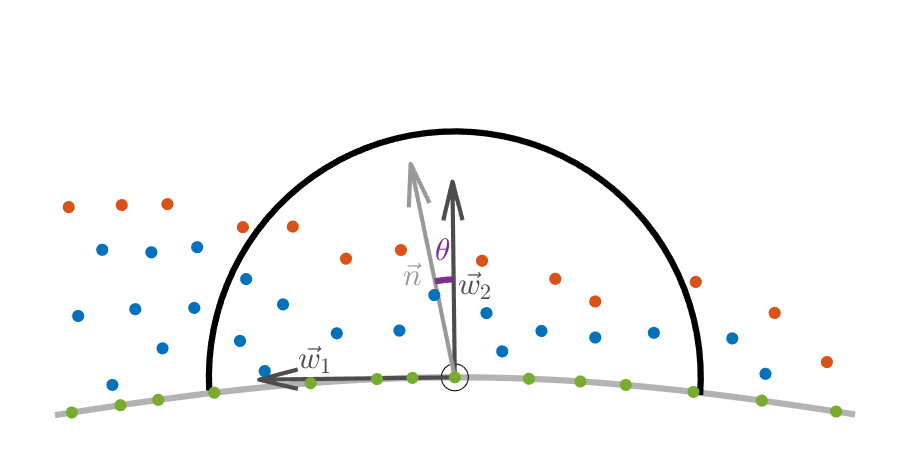}
  \label{Fig:EigenVector_b}}\\
  \subfloat[One eigenvalue (corresponding to eigenvector $\vec{w}_1$) is much larger than the other eigenvalue (corresponding to eigenvector $\vec{w}_2$). However, since the eigendirection $\vec{w}_2$ is ``far" away from the normal $\vec{n}$, this particle will not be transitioned to the thin film model.]{%
  \includegraphics[align=c, width=0.48\textwidth]{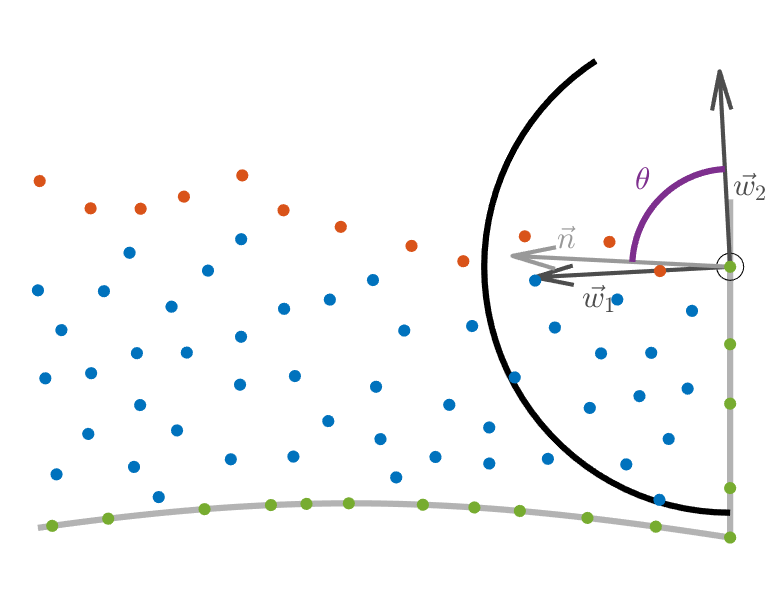}
  \label{Fig:EigenVector_c}}
	\caption{PCA eigenvalue and eigenvector criteria for detecting bulk to thin film transition. Wall particles are marked in green, interior particles in blue, and free surface particles in orange. The wall particle at which the bulk to thin film transition criterion is being evaluated is marked with an additional black circle, with its neighbourhood highlighted, and normal $\vec{n}$ marked. The wall is marked in grey, and eigenvectors of the local variance matrix are marked as $\vec{w}_k$. Note that throughout this work a 3D bulk model and 2D thin film model is considered. This figure shows a lower dimensional schematic for the ease of visualization.}
  \label{Fig:PCAcriterion}%
\end{figure}

In addition to the relative sizes of the eigenvalues, the direction of the corresponding eigenvectors also need to be examined. For a 3D particle $i$, let $\vec{w}_k$ for $k=1,2,3$ be the normalized eigenvectors with unit length corresponding to the eigenvalues $\lambda_k$. Consider the scenario described above of a significantly smaller $\lambda_3$ compared to $\lambda_1$ and $\lambda_2$. Here, $\vec{w}_1$ and $\vec{w}_2$ form the principal components or directions of largest variation (of the particle locations) in the local neighbourhood of particle $i$. Furthermore, $\vec{w}_3$ is the orthogonal direction where the particle locations vary the least in the neighbourhood. For a change of model from 3D to 2D, we must ensure that the direction $\vec{w}_3$ corresponds to the direction of depth of the resultant thin film, see Figure~\ref{Fig:EigenVector_b}. In certain geometric scenarios, it might be possible to have a significantly smaller $\lambda_3$ due to the boundary configuration. For example, consider a particle on a fixed wall right beside a free surface, as shown in Figure~\ref{Fig:EigenVector_c}. Here the small eigendirection points outside the domain, and does not indicate the presence of a thin fluid film. 

This scenario is accounted for by imposing a condition on the direction of $\vec{w}_3$. Since the height of a fluid film at a location is, by definition, the extension of the film perpendicular to the surface, I check the direction of $\vec{w}_3$ in relation to the surface normal $\vec{n}$ at the particle. To account for a surface normal in either direction, I check the angular distance of $\vec{w}_3$ with both $\vec{n}_i$ and $-\vec{n}_i$. Thus, I impose the restriction of 
\begin{equation}
\label{Eq:EigenvectorLimit}
    | \vec{n}_i \cdot \vec{w}_3 | < \cos(\theta^*) \,,    
\end{equation}
for a particle to be considered for transition to the thin film model. Here, $\theta^*$ is the threshold of maximum deviation allowed. Through numerical trial and error, I set a maximum deviation of the two vectors to be $\theta^* = \frac{\pi}{6}$.  




~\\

In addition, to these criteria, I allow for user-defined criteria for model transition. An example of such a situation is shown with the numerical results in Section~\ref{sec:WaterCrossing}. The overall check of 3D to 2D model transition is summarized in Algorithm~\ref{alg:3Dto2Dcheck}

\begin{algorithm}
    \caption{Detecting model transition: 3D to 2D} \label{alg:3Dto2Dcheck}
    \begin{algorithmic}
    \State{\textbf{Output:} Flagged particles for bulk to thin film transition}
    \Statex
    \Function{Loop}{$i$: all 3D wall particles}
        \State TransitionFlag(i) $\gets$ false \Comment{Not flagged for model transition}
        \If{Resolution criterion}    \Comment{See Section~\ref{sec:Resolution_3Dto2D}}
            \If{Eingenvalue and eigenvector criteria}    
            \Comment{Eq.\,\eqref{Eq:EigenvalueLimit}, Eq.\,\eqref{Eq:EigenvectorLimit}}
                 \State TransitionFlag(i) $\gets$ true
            \EndIf
            \If{User-defined criteria}    
                 \State TransitionFlag(i) $\gets$ true
            \EndIf      
        \EndIf
    \EndFunction
    \end{algorithmic}
\end{algorithm}


\subsection{Detecting 2D to 3D transition}
\label{sec:FilmToBulk}

Detecting model transition from the 2D thin film model to the 3D bulk flow models follows a similar, but simplified, procedure as the converse direction explained above in Section~\ref{sec:BulkToFilm}. Consider a 2D particle $i$. Note that by definition of the 2D model, all thin film particles are wall particles. 

The primary criterion for this transition is the resolution. If the user-prescribed resolution of the \emph{3D model} at the location $\vec{x}_i$ is deemed sufficient for the use of a 3D model, then the particle is flagged for model transition. This is quantified by comparing the 3D resolution with the computed film height at that location. Specifically, if the following condition holds, the particle $i$ is flagged for model transition from 2D to 3D
\begin{equation}
\label{Eq:2Dto3Dcond}
    H_{\text{film}}(\vec{x}_i) \geq h_{\text{3D}}(\vec{x}_i) \,.
\end{equation}
\begin{remark}
    At a given location $\vec{x}_0(t)$, the user can prescribe different resolutions for the thin-film and bulk flow models. $h_{\text{3D}}(\vec{x}_i)$ denotes the resolution of the bulk model at the location of particle $i$, which might be different from the thin film resolution of particle $i$.
\end{remark}

Consider the scenario introduced in Section~\ref{sec:Intro} where fluid is collecting at a location $\vec{x}_j \in \Omega^{2D}$. The computed film height at that location $H_{\text{film}}(\vec{x}_j)$ keeps increasing as more fluid collects. Once Eq.\,\eqref{Eq:2Dto3Dcond} holds, the model applied at $\vec{x}_j$ will be changed to the 3D bulk flow model, as desired.

Furthermore, additional user-defined criterion are also allowed, and are showcased in Section~\ref{sec:WaterCrossing}.


\subsection{Space and time smoothing}

For model transition in either direction, in addition to the criteria defined above, I impose two further restrictions to prevent numerical instabilities.
\begin{enumerate}
    \item If a single particle is flagged to undergo model transition, with no other particle in its neighbourhood being of the target model, model transition is not done. This prevents the scenario where a particle has no neighbours governed by the same model as itself.  
    \item Model transition is not performed for particles that have undergone a model transition in the previous time-step.
\end{enumerate}
These restrictions impose a notion of smoothness on the model transition. Empirically, I observe that this helps maintain stability of the simulations.


\section{Performing model transition}
\label{sec:Transition}

\begin{figure}
  \centering
  \subfloat[Transition from bulk particles to thin film particles.\\
  Left: Wall particles identified for model transition highlighted. \\
  Middle: Interior and free surface particles to be deleted identified and mapped to a particle undergoing model transition.\\
  Right: Particles after model transition.]{%
  \includegraphics[align=c,width=0.3\textwidth]{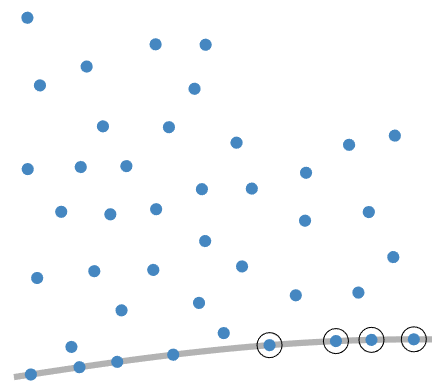}
  \hspace{0.02\textwidth}
  \includegraphics[align=c,width=0.3\textwidth]{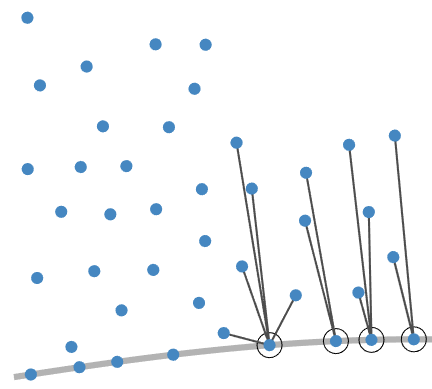}
  \hspace{0.02\textwidth}
  \includegraphics[align=c,width=0.3\textwidth]{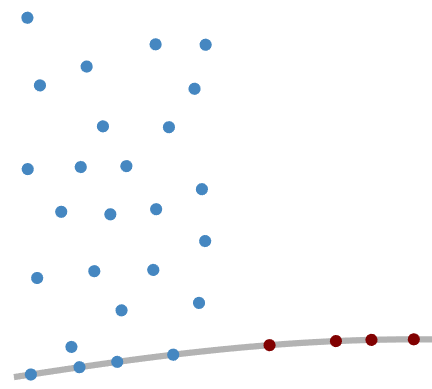} 
  \label{Fig:Transition_3Dto2D}} \\
  \subfloat[Transition from thin film particles to bulk particles.\\
  Left: Particles identified for model transition highlighted. \\
  Middle: Flagged particles transitioned to the bulk model.\\
  Right: New bulk particles added.]{%
  \includegraphics[align=c,width=0.3\textwidth]{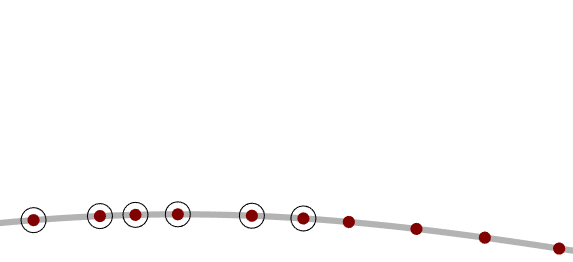}
  \hspace{0.02\textwidth}
  \includegraphics[align=c,width=0.3\textwidth]{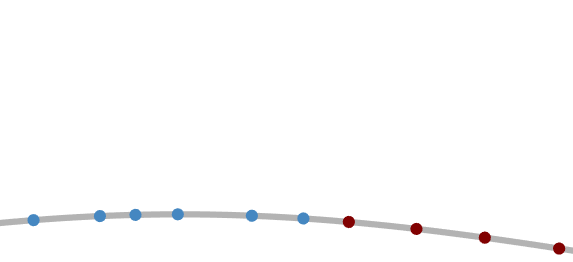}
  \hspace{0.02\textwidth}
  \includegraphics[align=c,width=0.3\textwidth]{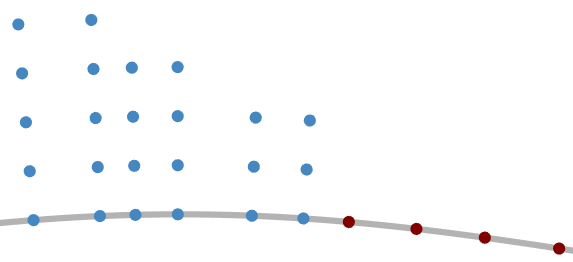} 
  \label{Fig:Transition_2Dto3D}} 
	\caption{Performing model transition from the bulk model to the thin film model (figure a), and from the thin film model to the bulk model (figure b). Bulk particles are shown in blue, and thin film particles in maroon. Note that throughout this work a 3D bulk model and 2D thin film model is considered. This figure shows a lower dimensional schematic for the ease of visualization.}
  \label{Fig:ModelTransition}%
\end{figure}

Section~\ref{sec:Where} dealt with detecting when a particle needs to be transitioned from one model to another. Once these checks are complete, all particles that need to undergo model transition are flagged. This section deals with how model transition is done for these flagged particles. There are two major challenges that need to be addressed to perform this transition:
\begin{enumerate}
    \item[I] \emph{Discretization adaptivity:} As mentioned earlier, the 3D and 2D models have different types of discretization. In the 3D flow model, the entire domain is discretized with particles. In contrast, in the 2D thin film model, particles only discretize the surface over which the fluid flows. The presence of the fluid in the direction normal to the surface is represented by the height function, see Eq.\,\eqref{Eq:Height}. Thus, as the model is being transitioned from one to another, the type of discretization also needs to be changed. 
    \item[II] \emph{Data transfer between models:} This second important challenge to address is how to accurately transfer system variables from one model to another during model transition. 
\end{enumerate}
Similar to Section~\ref{sec:Where}, I address these issues separately for model transition from 3D to 2D in Section~\ref{sec:BulkToFilm_How}; and for 2D to 3D transition in Section~\ref{sec:FilmToBulk_How}. A schematic of the model transition in both directions is shown in Figure~\ref{Fig:ModelTransition}.

Efficient neighbour searching is a critical part of every meshfree method. See, for example,~\cite{awile2012fast, lu2023local, onderik2008efficient}. Existing work towards this directly applies to both the models considered here. In this work, I maintain separate data structures (search trees) for the 3D model and the 2D model. Thus, a particle undergoing model transition must be removed from the data structure of the original model, and added to the data structure storing the new model. For MPI-based distributed memory parallel simulations, I observe that maintaining separate search trees resulted in better load balancing. However, this could be code dependent. Optimal strategy for load balancing remains a topic that needs investigation. Ideas in this regard from literature on the interaction of a fluid with a thin or flexible structure could be carried over to the present situation (see, for example, \cite{mehl2016parallel}). 

%
%

%
%


\subsection{3D to 2D transition}
\label{sec:BulkToFilm_How}

Consider a 3D wall boundary particle $i$ flagged for transition to the 2D model. Since the 2D thin film model only requires a discretization on the surface, a subset $\mathcal{T}_i \subset S_i$ of interior and free surface neighbours of $i$ must be deleted. All system variables from the particles in $\mathcal{T}_i$ must be mapped to the updated 2D particle $i$. First, I deal with the question of how to determine $\mathcal{T}_i$, and then how data mapping will be done. 

\subsubsection*{Determining $\mathcal{T}_i$}~\\
To ensure accurate data mapping, every particle being deleted needs to be mapped to a particle undergoing transition. For instance, for mass conservation, the mass of each particle being deleted must be lumped onto a nearby wall particle undergoing transition. For this, I defined $\mathcal{T}_i$ to be the set of all interior or free surface particles whose closest wall particle neighbour is $i$. The union of all these sets $\cup_i \mathcal{T}_i$ are the particles that have to be deleted once data mapping is complete. See the middle image of Figure~\ref{Fig:Transition_3Dto2D}.

Computationally, this mapping is determined in a two-step procedure. First, for each 3D particle $i$ flagged for model change to 2D, each of its interior and free surface neighbours are marked as candidates for deletion. Subsequently, consider a particle $b$ marked as a candidate for deletion, and search the neighbourhood $S_b$ for the nearest wall particle $j^w$. The superscript $w$ indicates that the particle is a wall boundary particle. If $j^w$ has been flagged for model transition, $b$ is added to $\mathcal{T}_{j^w}$. If particle $j^w$ has not been flagged for model transition, or if there is no wall particle in $S_b$, then particle $b$ is removed from the list of candidate particles to be deleted.


\subsubsection*{Data mapping}~\\
Before the particles in $\cup_i \mathcal{T}_i$ are deleted, their data first needs to be mapped onto the particles undergoing model change. Most physical properties of the particle $i$ in the 2D model are simply computed by an averaging of the data across the 3D model. For a system variable $\phi$, its updated value after data model transition is determined as
\begin{equation}
    \phi_i^{\text{new}} = \frac{ \phi_i + \sum_{j \in \mathcal{T}_i} \phi_j }{1 + n(\mathcal{T}_i)} \,,
\end{equation}
where $n(\mathcal{T}_i)$ is the number of particles in $\mathcal{T}_i$. This averaging of values is done for a majority of system variables: velocity $\vec{v}$, pressure $p$, density $\rho$, and viscosity $\eta$. Note that this averaging is also physically meaningful, since the 2D thin film flow model is obtained via a depth-averaging of the 3D bulk flow model \cite{bharadwaj2022discrete}. However, for mass, rather than averaging, a mass lumping is needed to ensure mass conservation
\begin{equation}
    m_i^{\text{new}} = \hat{m}_i + \sum_{j \in \mathcal{T}_i} \hat{m}_j  \,.
\end{equation}
Recall that for collocation based meshfree methods used for the bulk flow model, mass is not inherent to the numerical scheme. Rather, a representative notion of mass is prescribed for post-processing, which is used here. See \cite{suchde2023volume} for details. The notation $\hat{m}$ denotes the representative mass for the bulk collocation solver, while the notation $m$ denotes the actual mass for the mass-particle based thin film solver. The difference between these two methods are explained in detail in \cite{suchde2023volume}. 

Once the updated mass $m_i^{\text{new}}$ and density $\rho_i^{\text{new}}$ are known, the droplet diameter can be determined with the relation $m = \rho V$, which gives $d_i^3 = \frac{6 m_i^{\text{new}} }{ \pi \rho_i^{\text{new}} }$. The height $H_{\text{film}}$ of particle $i$ in the 2D thin film model is set as the maximum normal distance of $i$ from all particles in $\mathcal{T}_i$
\begin{equation}
    H_{\text{film}, i} = \max_{j \in \mathcal{T}_i} | (\vec{x}_j - \vec{x}_i) \cdot \vec{n}_i | \,,
\end{equation}
where $\vec{n}_i$ is the unit boundary normal of particle $i$.

~\\
The procedure of transitioning a 3D particle to a 2D one is summarised in Algorithm~\ref{alg:3Dto2D}, and illustrated in Figure~\ref{Fig:Transition_3Dto2D}.
\begin{algorithm}
    \caption{Performing model transition: 3D to 2D} \label{alg:3Dto2D}
    \begin{algorithmic}
    \State{\textbf{Input:} Flagged 3D particles that will undergo model transition}
    \Statex  
    \Function{Loop}{all 3D particles flagged for model transition to 2D}
        \State Mark all interior and free surface neighbours as candidates for deletion
    \EndFunction
    \Statex  
    \Function{Loop}{$b$: all 3D candidates for deletion}
        \State \textbf{find} nearest wall boundary particle $j^w \in S_b$
        \If{$j^w$ is undergoing transition to 2D}    
            \State Data mapping onto particle $j^w$
            \State Delete particle $b$
        \Else
            \State Remove $b$ from deletion candidate list \Comment{This particle will not be deleted}
        \EndIf   
    \EndFunction
    \Statex  
    \Function{Loop}{all 3D particles flagged for model transition to 2D}
        \State Remove particle from data structure and neighbours lists of 3D model
        \State Add particle to data structure and neighbours lists of 2D model
    \EndFunction
    \end{algorithmic}
\end{algorithm} 
%


\subsection{2D to 3D transition}
\label{sec:FilmToBulk_How}

While the transition from 3D to 2D in Section~\ref{sec:BulkToFilm_How} required the deletion of particles, the transition from 2D to 3D considered in this section requires the addition of new particles to create a bulk discretization. Consider a 2D particle $i$ flagged for transition to 3D, with film height $H_{\text{film}, i}$ and unit boundary normal $\vec{n}_i$ pointing inwards, i.e. in the direction of the fluid. For the change in discretization, I add particles along the direction $\vec{n}_i$ until a distance of $H_{\text{film}, i}$ (see the rightmost image of Figure~\ref{Fig:Transition_2Dto3D}). This ensures that the resultant depth of the bulk model matches that of the thin film model.

\subsubsection*{Adding particles}~\\
Recall from Section~\ref{sec:3Dmodel} that the inter-particle spacing in the bulk model varies between $r_{\text{min}} h$ and $r_{\text{max}} h$ where $h=h(\vec{x}, t)$ is the \emph{3D} resolution. To maintain these distances in the newly added particles, I assume 3D particles to be added at an approximate distance of
\begin{equation}
    \tilde{\psi} = \frac{r_{\text{min}} h + r_{\text{max}} h}{2} \,.
\end{equation}
As a result the number of particles added along $\vec{n}_i$ is given by
\begin{equation}
    \gamma_i = \left\lceil \frac{H_{\text{film}_i}}{\tilde{\psi}_i} \right\rceil \,,
\end{equation}
where $\lceil \cdot \rceil$ is the ceiling function. Thus, the actual distance between particles added is $\psi_i = \frac{H_{\text{film}, i}}{\gamma_i}$, and the location of each new particle is given by $\vec{x}_{i_k} = \vec{x}_i + k \psi_i \vec{n}_i$ for $k = 1,2,\hdots, \gamma_i$. The particle furthest away from $i$, obtained by $k=\gamma_i$ is set as a free surface particle. Once particle addition is complete for all particles flagged for model transition, a free surface check is run on all newly added particles. This can follow any free surface detection method \cite{liu2023high, lu2016finite, saucedo2019three, tiwari2003particle} in literature. 

\subsubsection*{Data mapping}~\\
Data mapping for the newly created particles is straightforward. The velocity, pressure, density and viscosity for a newly created particle $i_k$ is set to the same as that for particle $i$. The original mass of particle $i$ is split evenly between all the newly created particles $\hat{m}_i^{\text{new}} = \hat{m}_{i_k} = \frac{m_i}{1+ \gamma_i}$.

~\\
The procedure of transitioning a 2D particle to 3D one is summarised in Algorithm~\ref{alg:2Dto3D}, and illustrated in Figure~\ref{Fig:Transition_2Dto3D}.
\begin{algorithm}
    \caption{Performing model transition: 2D to 3D} \label{alg:2Dto3D}
    \begin{algorithmic}
    \State{\textbf{Input:} Flagged 2D particles that will undergo model transition}
    \Statex  
    \Function{Loop}{all 2D particles flagged for model transition to 3D}
        \State Data mapping for 3D model
        \State Remove particle from data structure and neighbours lists of 2D model
        \State Add particle to data structure and neighbours lists of 3D model
    \EndFunction
    \Statex  
    \Function{Loop}{all particles just transitioned to 3D}
        \State Perform addition/deletion on the wall \Comment{to match desired resolution of 3D model}
    \EndFunction
    \Statex  
    \Function{Loop}{all particles just transitioned to 3D}
        \State Create new bulk particles in the normal direction
        \State Data mapping for new particles        
        \State Set appropriate boundary type (interior/free surface) for new particles
    \EndFunction
    \end{algorithmic}
\end{algorithm} 
%


\section{Data communication}
\label{sec:DataComm}

\begin{figure}
  \centering
  \subfloat[Bulk particles (blue, left) and thin film particles (maroon, right) beside each other. Ghost particles need to be created for data communication between these particles.]{%
  \includegraphics[align=c,width=0.3\textwidth]{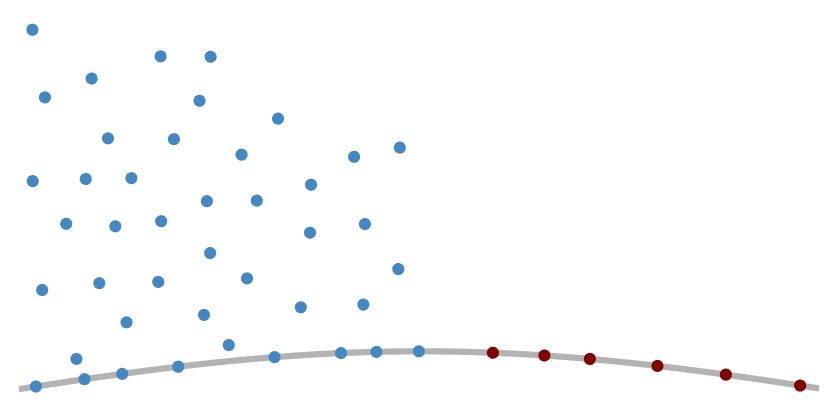}
  \label{Fig:Ghost_a}} \hspace{0.02\textwidth}
  \subfloat[Ghost particle sources: particles near particles of the other model are marked as sources for ghost particles.]{%
  \includegraphics[align=c,width=0.3\textwidth]{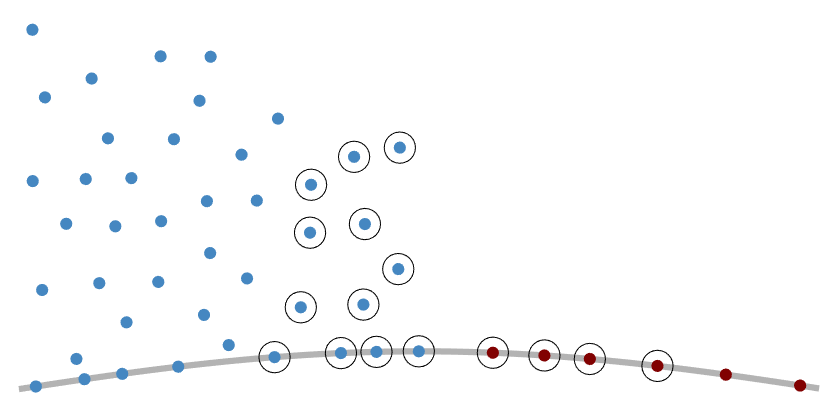}
  \label{Fig:Ghost_b}} \hspace{0.02\textwidth}
  \subfloat[Intermediate ghost particles: Each source particle creates a ghost particle at the same location with the same model.]{%
  \includegraphics[align=c,width=0.3\textwidth]{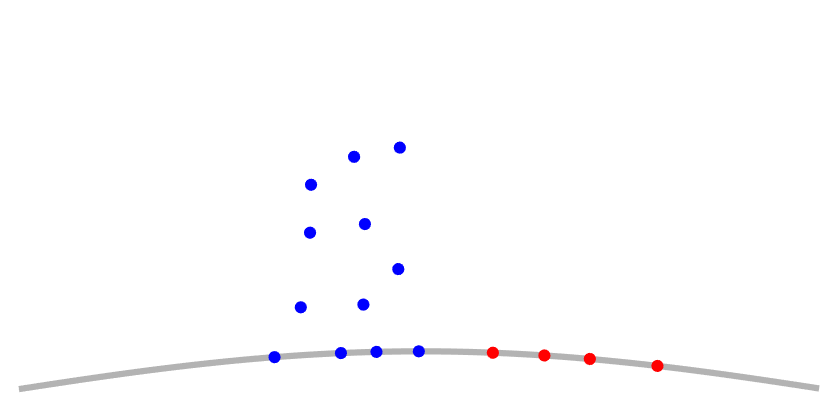}
  \label{Fig:Ghost_c}} \\
  \subfloat[Final ghost particles: After model transition on the set of intermediate ghost particles.]{%
  \includegraphics[align=c,width=0.3\textwidth]{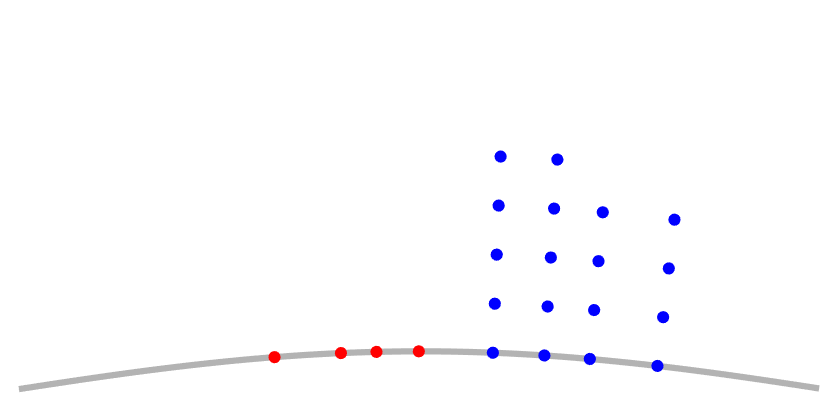}
  \label{Fig:Ghost_d}} \hspace{0.02\textwidth}
  \subfloat[All bulk particles: regular bulk particles (left), and ghost bulk particles (right, bright blue). ]{%
  \includegraphics[align=c,width=0.3\textwidth]{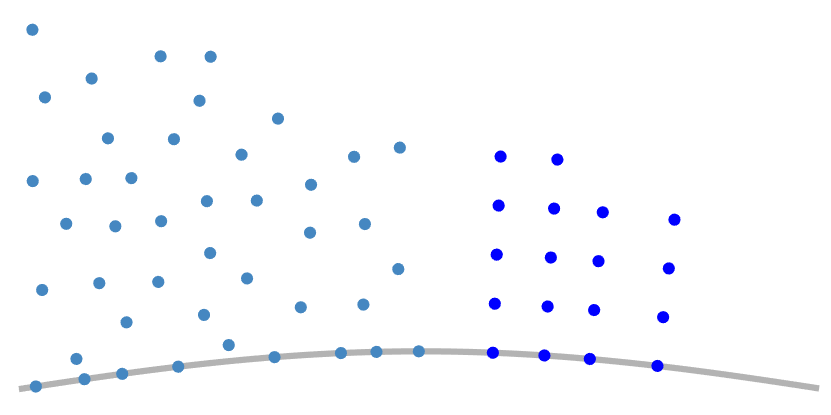}
  \label{Fig:Ghost_e}} \hspace{0.02\textwidth}
  \subfloat[All thin film particles: regular thin film particles (right), and ghost thin film particles (left, bright red). ]{%
  \includegraphics[align=c,width=0.3\textwidth]{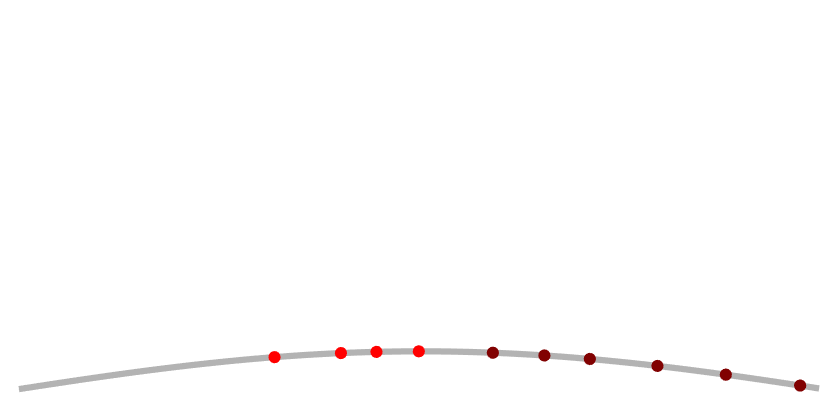}
  \label{Fig:Ghost_f}}
	\caption{Creation of ghost particles. Bulk particles are shown in blue, and thin film particles in maroon. Ghost particles are shown in a brighter shade for each model. Note that throughout this work a 3D bulk model and 2D thin film model is considered. This figure shows a lower dimensional schematic for the ease of visualization. A representation of the same in $\mathbb{R}^3$ is shown in the results section in Figure~\ref{Fig:AdvectionDomain}.}
  \label{Fig:GhostParticles}%
\end{figure}

The last two sections dealt with where and how to perform model transition at each time step. Once all needed model change is performed, the final step in the model adaptivity framework is to perform time integration of the respective models. The models are solved sequentially, with the 3D solver running first. At a time step when particles of both models are present, there will be particles of one model with neighbours of another model, see Figure~\ref{Fig:Ghost_a}. Since the two models have different discretization and different data stored, extra work needs to be done for data communication between the models. 

This same issue also occurs when two models are coupled without any adaptivity, as is done in existing literature. I adopt the usage of a so-called buffer zone from literature, also referred to as overlapping method \cite{ling2023two, pan2024variable}, handshake region or blending region \cite{budarapu2019multiscale, talebi2013molecular}. In this approach, in a small region centred at the interface between the models, a buffer zone exists where both models are solved. Data is communicated between models only through this buffer zone. In the present work, the buffer zone is populated with particles of each model. As a result, particles outside this buffer zone can be associated with neighbourhoods containing only particles of the same model (see Section~\ref{sec:GhostApproximations}), which significantly simplifies approximation procedures. Note that since the type of discretization is different for the 3D and 3D models considered here, the particles of the two models in the buffer region will have different locations. 

\begin{remark}
   The interface between two models is not stored explicitly nor implicitly through a level set or similar approach. In fact, the interface is not even computed. The term is used for a purely virtual interface across which a particle of a particular model interacts with a particle of the other model. 
\end{remark}

\subsection{Creating ghost particles}
\label{sec:GhostParticles}

The extra particles added in the buffer region are referred to as ghost particles, while non-ghost particles will be referred to as regular particles. The creation of ghost particles follows a three-step procedure. 
\begin{enumerate}
    \item In the first step, particles that will create ghost particles are flagged, see Figure~\ref{Fig:Ghost_b}. 
    \item Subsequently, for all flagged particles, a ghost particle is created at the same location, see Figure~\ref{Fig:Ghost_c}. 
    \item In the final step, model transition is performed on these ghost particles, see Figure~\ref{Fig:Ghost_d}.
\end{enumerate}

For the first step, consider a particle $i$ and all nearby particles within a distance of $h$ from $i$. If there exists a nearby particle $j$ which is governed by a different model than $i$, then both $i$ and $j$ are flagged to create ghost particles. Thus, if $i$ is a 3D particle, all neighbouring 2D particles are flagged, and vice versa. In the second step, for each flagged particle $j$, a ghost particle is created at the same location as $j$ with the same model as $j$. This forms a set of intermediate ghost particles. Finally, model transition is performed on all intermediate ghost particles, as done in Section~\ref{sec:Transition}, to get the final set of ghost particles. This procedure is illustrated in Figure~\ref{Fig:GhostParticles}.

To further describe this procedure, consider a 3D particle $i$. The creation of ghost particles near $i$ starts by flagging all 2D particles within a distance of $h$ from $i$. Each of these flagged 2D particles create co-incident 2D ghost particles. Finally, all 2D ghost particles are transitioned to 3D ghost particles, by adding further 3D ghost particles in the normal direction and appropriate data mapping, as per Section~\ref{sec:FilmToBulk_How}. These newly created 3D ghost particles contain information from the 2D model in the buffer region, and will be used in all approximations performed at particle $i$. Similarly, for a 2D particle $i$, all 3D neighbours are flagged as sources for ghost particles. The flagged 3D particles create 3D ghost particles, which are finally transitioned to 2D with the discretization adaptivity and data mapping procedure laid down in Section~\ref{sec:BulkToFilm_How}. This includes the deletion of all interior and free surface ghost particles, with their data mapped to the corresponding wall boundary ghost particles before its model is changed to the 2D thin film model.



%

In typical model coupling in existing literature, two models are solved beside each other and the interface is either fixed or can be tracked. In contrast, here, the interface between models is completely dynamic. Thus, the buffer region possibly needs to be updated at every time step. For a general scenario, where the interface is dynamic, I delete old ghost particles at the start of the time step, and then recreate new ghost particles after the model transition is complete. For the special case where the interface is fixed, ghost particles from the previous time step are not deleted and only their data is updated every time step. 

The procedure of creating ghost particles and the buffer zone is summarised in Algorithm~\ref{alg:Ghost}, and visualized in Figure~\ref{Fig:GhostParticles}. Note that since ghost particles are created for the 2D and 3D models, this approach forms a two-way coupling of the models. 
\begin{algorithm}
    \caption{Creating ghost particles} \label{alg:Ghost}
    \begin{algorithmic}
    \Function{Loop}{$i$: all particles} \Comment{across both models}
        \State Flag all other model particles within a distance of $h$ \Comment{see Figure~\ref{Fig:Ghost_b}}
    \EndFunction
    \Statex  
    \Function{Loop}{$j$: all flagged particles} \Comment{across both models}
        \State Create ghost particle of the same model as $j$ at location $\vec{x}_j$  \Comment{see Figure~\ref{Fig:Ghost_c}}
    \EndFunction
    \Statex  
    \Function{Loop}{all newly created ghost particles}
        \State Perform model transition \Comment{according to algorithms \ref{alg:3Dto2D} and \ref{alg:2Dto3D}}  \\
        \Comment{see Figure~\ref{Fig:Ghost_d}}
    \EndFunction
    \end{algorithmic}
\end{algorithm} 

The size of the buffer region is often considered a parameter in methods that couple two models across a fixed interface. In the present work, this is fixed to the interaction radius $h$ at that location. Since all computations are local, based on neighbouring particles within the interaction radius, increasing the size of the buffer region beyond $h$ will not increase accuracy. 

\subsection{Derivatives, approximations and time integration}
\label{sec:GhostApproximations}

Consider a particle $i$, with a set of neighbouring particles $S_i$. This set of neighbouring particles $S_i$ only consists of neighbours being governed by the same model as $i$, and does not include ghost particles. Furthermore, consider the set $S_i^g$ of ghost particle neighbours of $i$. Once again, $S_i^g$ only includes ghost particles of the same model as $i$. The set $S_i^g$ will be empty if particle $i$ is sufficiently away from all particles of the other model. Let the complete neighbouring set, consisting of both regular and ghost neighbours, be represented as $\overline{S}_i = S_i \cup S_i^g$.  

\begin{remark}
    The free surface check for newly added particles in Section~\ref{sec:FilmToBulk_How} must include ghost particle neighbours as well. 
\end{remark}

Before addressing derivative computation in the presence of ghost particles, I first recall the regular computation of derivatives in a GFDM approach without ghost particles. Consider a discrete function $u$ defined at each particle location. The derivatives of $u$ at particle $i$ are approximated as
\begin{equation}
	\label{Eq:GFDM_Definition}
	\partial^* u(\vec x_i)\approx \partial^*_i u = \sum_{j\in S_i}c_{ij}^* u_j \,,
\end{equation}
where ${}^*=x,y, z, \Delta$ represents the differential operator being approximated, $\partial^*$ represents the continuous $^*$-derivative, and $\partial^*_i$ represents the discrete derivative at particle $i$. The stencil coefficients $c_{ij}^*$ are found either using Taylor expansions, or equivalently by ensuring that monomials up to the desired order are exactly differentiated \cite{jacquemin2020taylor, suchde2018meshfree}. 

In the presence of ghost particles, if $S_i^g \neq \varnothing$, all derivatives and approximations are computed using the extended neighbourhood $\overline{S}_i$, including both regular and ghost neighbours. Thus, we have
\begin{align}
    \partial^*_i u &= \sum_{j\in \overline{S}_i}c_{ij}^* u_j \,, \\ 
    &= \sum_{j\in S_i}c_{ij}^* u_j + \sum_{j\in S_i^g}c_{ij}^* u_j \,. \label{Eq:GhostDerivatives}
\end{align}

Now, consider integrating from time level $t^{(n)}$ to  $t^{(n+1)}$, where the bracketed superscript denotes the time level. Explicit time integration is straight-forward, by replacing $u_j$ in Eq.\,\eqref{Eq:GhostDerivatives} by $u_j^{(n)}$. 

As explained above, due to the dynamic nature of the buffer region, the ghost particles are recreated at every time step. As a result, there is no need to update the system variables at the ghost particles. Thus. derivatives are never computed at ghost particles; and all large sparse implicit systems only contain regular particles, and no ghost particles. In the special cases when the same buffer region exists for multiple time steps, rather than recreating the ghost particles, the data at the ghost particles are recomputed as explained in Section~\ref{sec:GhostParticles}. Since the data at ghost particles are not updated, derivative computation during implicit time integration is done using the unknown new time step value for regular neighbours, with the current time step value for ghost neighbours
\begin{equation}
    \partial^*_i u = \sum_{j\in S_i}c_{ij}^* u_j^{(n+1)} + \sum_{j\in S_i^g}c_{ij}^* u_j^{(n)} \,.
\end{equation}
The implicit terms in the first summation will be a part of the sparse system matrix, while the explicit terms of the second summation will be a part of the right-hand side. 


\section{Numerical Results}
\label{sec:Results}

The bulk flow and thin film models used here, and their implementations, have been introduced, verified and validated in the author's earlier work \cite{bharadwaj2022discrete, suchde2017flux, suchde2018meshfree}. Furthermore, the novelty in the present work lies in the adaptive combination of the models, and not the individual models themselves. Thus, this results section only focuses on the combination of the models, without any results with either of the models individually.

I consider test cases of increasing complexity. The first two cases test different parts of the introduced framework, while the third case considers the full scheme. The last test case considers the full scheme with added user-defined criteria for choosing the regions of applicability of the different models. A quick summary of the all the numerical simulations considered, and the motivation behind them, is described in Table~\ref{tab:Testcases}. Note that all test cases are in $\mathbb{R}^3$ with a 3D bulk flow model and a 2D thin film model.

\begin{table}
	\caption{Numerical experiments used to highlight the effectiveness of the introduced model adaptive framework, and the motivation behind each case.}
	\centering
	\label{tab:Testcases}
	{  
    \begin{tabular}{|p{0.02\textwidth}|p{0.35\textwidth}|p{0.55\textwidth}|}
	\hline
    & Test case &  Comments \\ 
	\hline \hline
1& Steady fluid in a shallow cylinder              &  Model transition without flow \newline
                                        Tests the model transition, and data mapping algorithms  \\\hline  
2& Advection diffusion in uniform flow          & Tests the model transition, and data mapping algorithms \newline
                                    Tests data communication using ghost particles \newline
                                        Convergence tests \\\hline
3& Cleaning jet          &  Full adaptive model \\\hline 
4& Automotive water crossing  &  Full adaptive model with additional user-defined criteria \newline
                                        Industrial test case\newline
                                        Simulation time comparison \\\hline
	\end{tabular}}
\end{table}
%

\subsection{Steady fluid in a shallow cylinder}
\label{sec:SteadyFlow}

As a first test case, I run the model transition algorithm with the PDE solvers switched off. Consider a cylinder of unit radius, filled partially with a stationary fluid. The resolution of the domain is varied such that it triggers the model transition criteria in the entire domain from 3D to 2D, and then vice versa.
Since the entire domain is solved either with a 3D model, or a 2D model, ghost particles and buffer zones do not play a role here. This case only tests the model transition detection, the discretization change during the model transition, and the data mapping between the models during model transition. The fluid considered is water, with density $\rho = 1000 \text{ kg/m}^3$ and viscosity $\eta = 10^{-3} \text{ Pa s}$.



\subsubsection*{3D to 2D}
\begin{figure}
  \centering
  \includegraphics[align=c,width=0.44\textwidth]{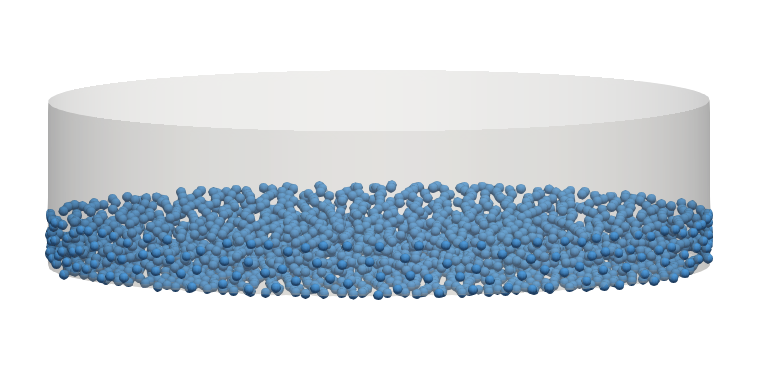}   
  \includegraphics[align=c,width=0.1\textwidth]{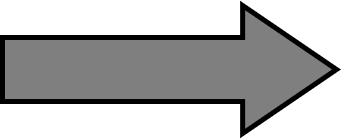} 
  \includegraphics[align=c,width=0.44\textwidth]{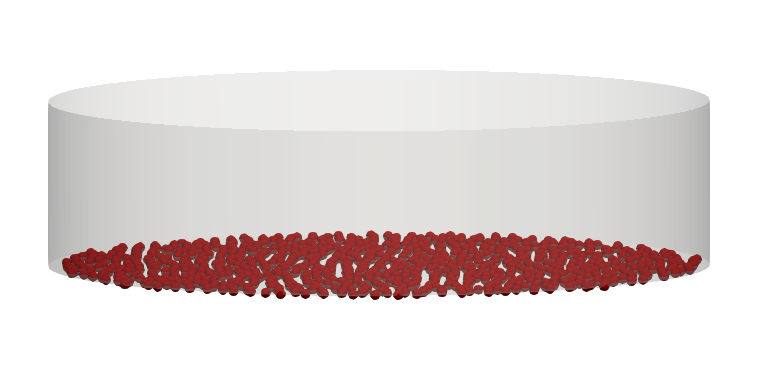}   
	\caption{Steady fluid in a shallow cylinder: Transition of a 3D model (left) to a 2D model (right).}
  \label{Fig:BulkToThinFilm}%
\end{figure}

~\\
Consider the cylinder filled to a height of $0.15 \text{m}$, and initialized with a 3D discretization, see Figure~\ref{Fig:BulkToThinFilm}. A resolution of $h=0.2$ is used for the initial discretization, which results in $N=4758$ particles in the domain. As the simulation starts, the resolution of the bulk model is changed to $h=0.3$. This results in the resolution becoming too coarse to capture the layer of fluid. I observe that the PCA criterion from Section~\ref{sec:PCA1} gets triggered at all wall boundary particles, resulting in a model transition to the 2D model. Thus, all particles transition to 2D at the same time. Figure~\ref{Fig:BulkToThinFilm} shows the discretization after the model transition is complete. The figure shows that all interior and free surface particles are deleted, as desired. Furthermore, the boundary particles on the vertical walls of the cylinder do not transition to a thin film model, since they  fail the eigenvector criterion of Section~\ref{sec:PCA2}. These wall particles are subsequently deleted during the transition process of the wall particles on the base of the cylinder.  

Figure~\ref{Fig:BulkToThinFilm} shows the discretization of the domain before and after transition. To verify the data transfer between the models, I check the following. The velocity of the fluid is completely at rest before and after model transition. The total mass and volume are the same up to numerical precision for both models. The analytically expected volume to be occupied by the domain is $0.15\pi \approx 0.4712$. Numerically, the volume occupied by the 3D discretization is $\sum_i {V}_i = 0.4631$. The slight discrepancy from the analytical volume (under $2 \%$) is expected due to the particle based discretization, see \cite{suchde2023volume} for more details. As expected from the data mapping algorithm, the total numerical volume occupied by the 2D model after transition matches this to numerical precision. Similarly, the total mass in the two models match up to numerical precision, $\sum_i {m}_i = 0.4631 \times 10^3 $. This verifies that the mapping from the interior and free surface particles being deleted to the wall particles works as intended, and that the conservative data transfer during 3D to 2D discretization works correctly.

\subsubsection*{2D to 3D}
\begin{figure}
  \centering
  \includegraphics[align=c,width=0.44\textwidth]{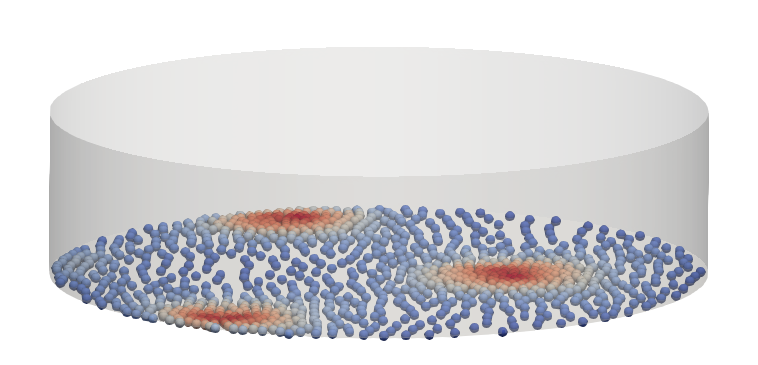}
  \includegraphics[align=c,width=0.1\textwidth]{./Figures/Arrow} 
  \includegraphics[align=c,width=0.44\textwidth]{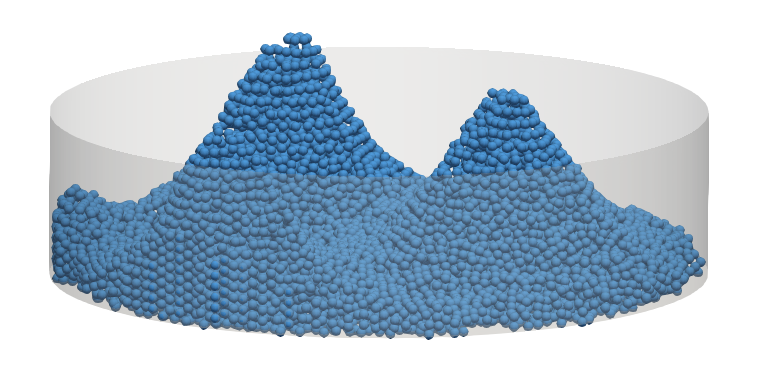}   
	\caption{Steady fluid in a shallow cylinder: Transition of a 2D model (left) to a 3D model (right). In the 2D figure (left), the colour indicates the computed height function. A comparison of the computed height function in the 2D model and the free surface in the 3D model is shown in Figure~\ref{Fig:ThinFilmToBulk_FS}.}
  \label{Fig:ThinFilmToBulk}%
\end{figure}

~\\
Now consider the same problem with the model transition happening in the other direction. The domain is initialized with a 2D discretization. Unlike the previous case, to increase the complexity of the problem, I initialize the film with a spatially varying thickness, as shown in Figure~\ref{Fig:ThinFilmToBulk}. The resolution of the 3D model is reduced throughout the domain to trigger the model transition criterion (Section~\ref{sec:FilmToBulk}) throughout the domain. Once again, I observe that the entire domain transitions to the 3D model at the same time. 

\begin{figure}
  \centering
  \includegraphics[align=c,width=0.7\textwidth]{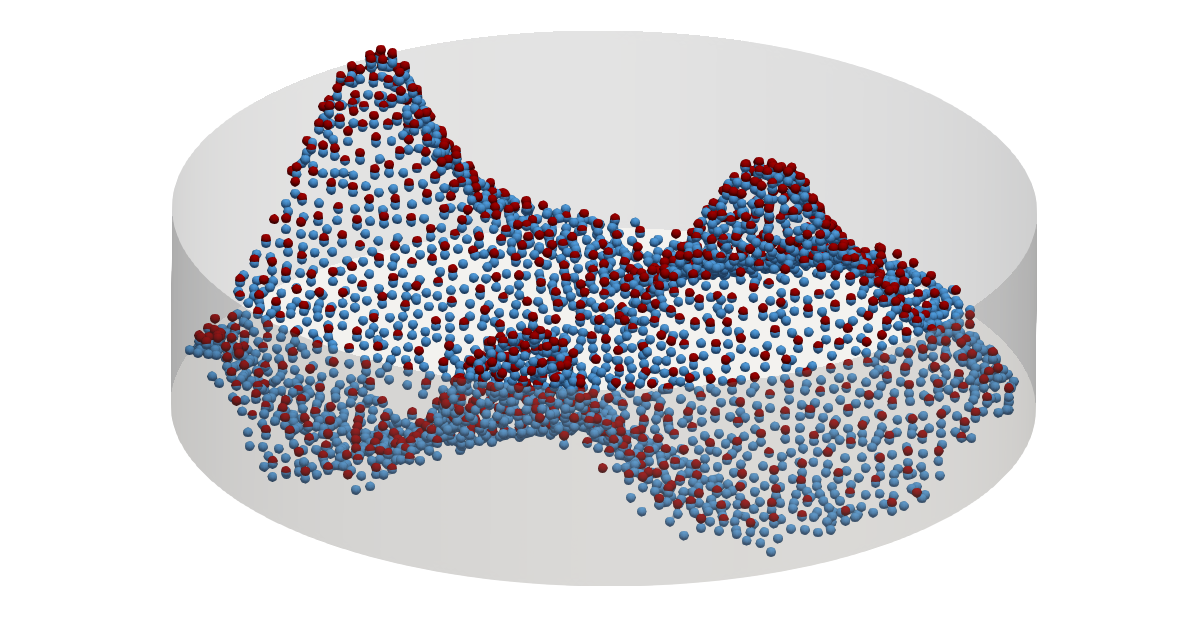}  
	\caption{Steady fluid in a shallow cylinder: Transition of the 2D model to the 3D model. Comparison of the computed height profile in the 2D model (maroon) and the free surface of the 3D model (blue). The individual domain discretization of each model are shown in Figure~\ref{Fig:ThinFilmToBulk}.}
  \label{Fig:ThinFilmToBulk_FS}%
\end{figure}

Figure~\ref{Fig:ThinFilmToBulk} shows the discretization of the domain before and after transition, while Figure~\ref{Fig:ThinFilmToBulk_FS} shows the comparison of the free surface profiles in each model. This considers the top most free surface profile in the 3D model (maroon) and the computed height function in the 2D model (blue) overlaid on each other. The figure shows that both models produce the same free surface profile. Since the PDE solvers are switched off in this test case, the spatially varying height of the fluid does not result in a tangential velocity. The fluid maintains the initially prescribed $0$ velocity, both before and after model transition. Once again, the total mass and volume are the same up to numerical precision for both models, with the total numerical volume $\sum_i {V}_i = 0.5592$, and the total numerical mass $\sum_i {m}_i = 0.5592 \times 10^3$. This highlights the conservative nature of the data mapping during model transition.
%

\subsection{Advection diffusion in uniform flow}
\label{sec:AdvectionDiffusion}

The previous test case checked the model transition detection, and the data transfer mechanism during model transition. Unlike the previous test case, here the full models are considered, with the PDEs switched on for both models. In addition, this test case also tests the data communication when both models are present beside each other with the help of ghost particles in a buffer zone. Consider inviscid flow of a fluid over a flat plate in the absence of gravity. A thin layer of fluid of uniform thickness is considered, with slip velocity boundary conditions on the plate.  

\begin{figure}
  \centering
  \subfloat[$t=0$. Only 2D discretisation present]{%
  \includegraphics[align=c,width=0.99\textwidth]{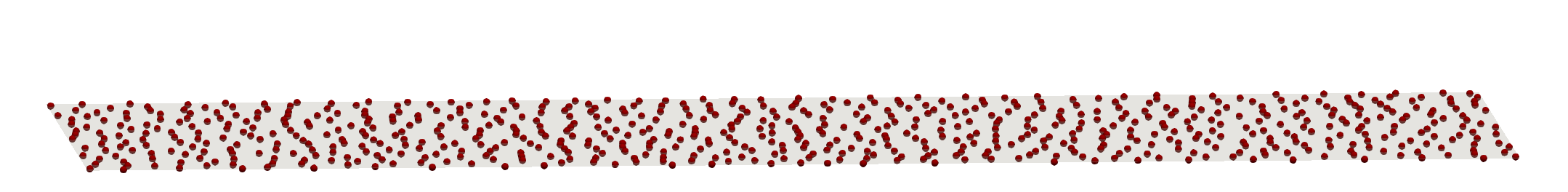}
  \label{Fig:AdvectionDomain_a}}\\
  \subfloat[$t=0.145$. 2D particles in maroon, 3D particles in blue.]{%
  \includegraphics[align=c,width=0.99\textwidth]{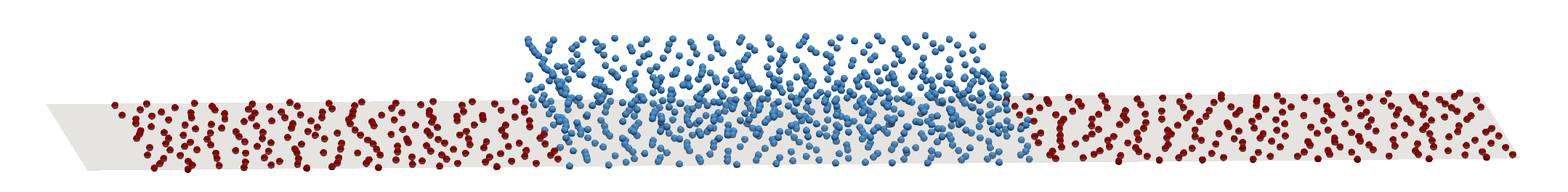}
  \label{Fig:AdvectionDomain_b}} \\ 
  \subfloat[$t=0.145$. Ghost particles only. 2D ghost particles in red, 3D ghost particles in blue.]{%
  \includegraphics[align=c,width=0.99\textwidth]{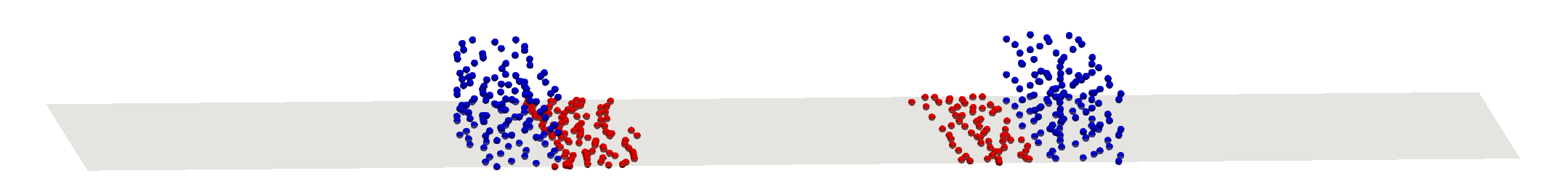}
  \label{Fig:AdvectionDomain_c}} \\
  \subfloat[$t=1.5$. 2D particles in maroon, 3D particles in blue.]{%
  \includegraphics[align=c,width=0.99\textwidth]{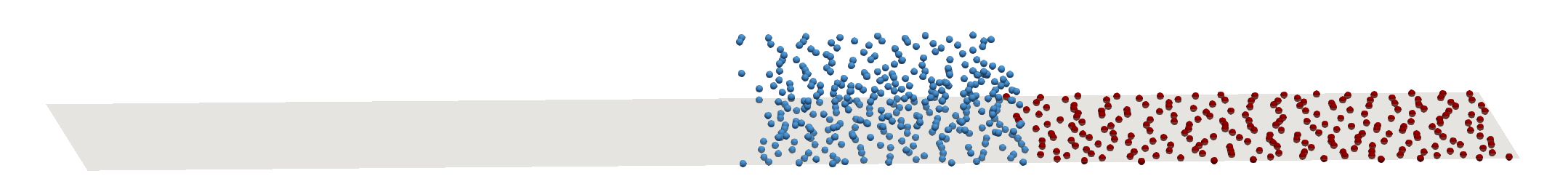}
  \label{Fig:AdvectionDomain_d}} 
	\caption{Advection diffusion in uniform flow: Domain discretisation at different times for $h=0.2$. The 2D particles are shown in maroon, and the 3D particles are shown in blue. Ghost particles of each discretisation type (Figure c) are shown in a brighter shade to distinguish them from regular particles. Ghost particles are not shown in figures a,b and d. The fluid is moving from left to right in a Lagrangian framework, without any inflow. Thus, there is no fluid on the left of the domain as the simulation progresses.}
  \label{Fig:AdvectionDomain}%
\end{figure}

Consider a long plate, with fluid flowing over the surface $[0,3] \times [0,1] \times [0]$ in the $xy$ plane. The fluid is discretized alternatingly with the 2D and 3D models, as shown in Figure~\ref{Fig:AdvectionDomain}. An initial velocity of $\vec{v} = (1,0,0)^T$ is prescribed throughout the domain, which corresponds to uniform flow from left to right in Figure~\ref{Fig:AdvectionDomain}. Since the flow is inviscid, and without gravity, both models produce a constant velocity field equal to the initial velocity. An additional advection diffusion equation in a Lagrangian framework is solved throughout the domain in both solvers. 
\begin{equation}
\label{Eq:AdvDiff}
    \frac{D T}{Dt} = \kappa \Delta T \,,
\end{equation}
for diffusivity $\kappa$, material derivative $\frac{D }{Dt} = \frac{\partial }{\partial t}+ \vec{v}\cdot \nabla$ for fluid velocity $\vec{v}$, and a temperature field $T$ being transported. Eq.\,\eqref{Eq:AdvDiff} is discretized using a Lagrangian movement step followed by an explicit Euler time integration for the diffusion term, and a GFDM type discretization for the Laplacian operator, as explained above. This test case is split into two parts: first with only advection $\kappa = 0$, followed by a convergence study for advection-diffusion with $\kappa \neq 0$. Both parts consider model transition in both directions: from 2D to 3D and 3D to 2D. 

\subsubsection*{Pure advection}~\\
Consider a pure advection case with $\kappa = 0$ in Eq.\,\eqref{Eq:AdvDiff}, with the domain and initial velocity as mentioned above. Initially, the entire domain follows the 2D thin film model, see Figure~\ref{Fig:AdvectionDomain_a}. From the first time step onwards, a 3D discretization and model is used in $x \in [1,2]$, see Figure~\ref{Fig:AdvectionDomain_b}. This corresponds to the middle third of the domain in $x$ direction, as shown in Figure~\ref{Fig:AdvectionDomain}. This implies two fixed interfaces between the 2D and 3D models. However, the use of a Lagrangian framework means that particles entering $x \in [1,2]$ are transitioned to the 3D model, while particles leaving this region are transitioned to the 2D model in every time step. Thus, the ghost particles in the buffer zone (see Figure~\ref{Fig:AdvectionDomain_c}) are also recreated at every time step. 

\begin{figure}
  \centering
  \subfloat[$t=0$]{     \includegraphics[align=c,width=0.9\textwidth]{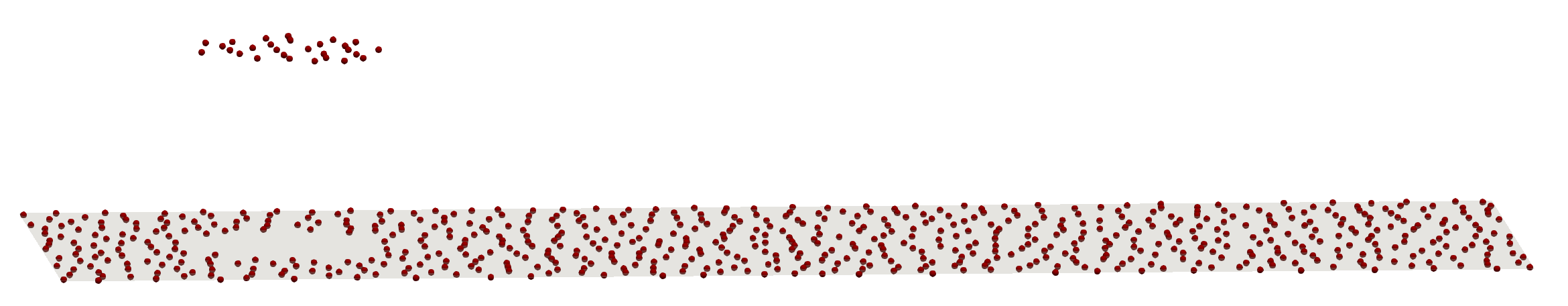}  \label{Fig:ARa}}\\
  \subfloat[$t=0.445$]{ \includegraphics[align=c,width=0.9\textwidth]{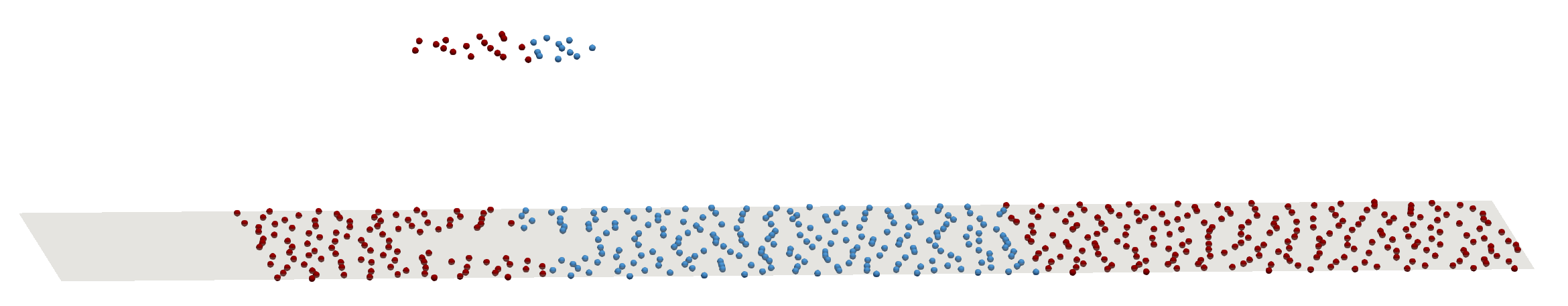}  \label{Fig:ARb}}\\
  \subfloat[$t=0.895$]{ \includegraphics[align=c,width=0.9\textwidth]{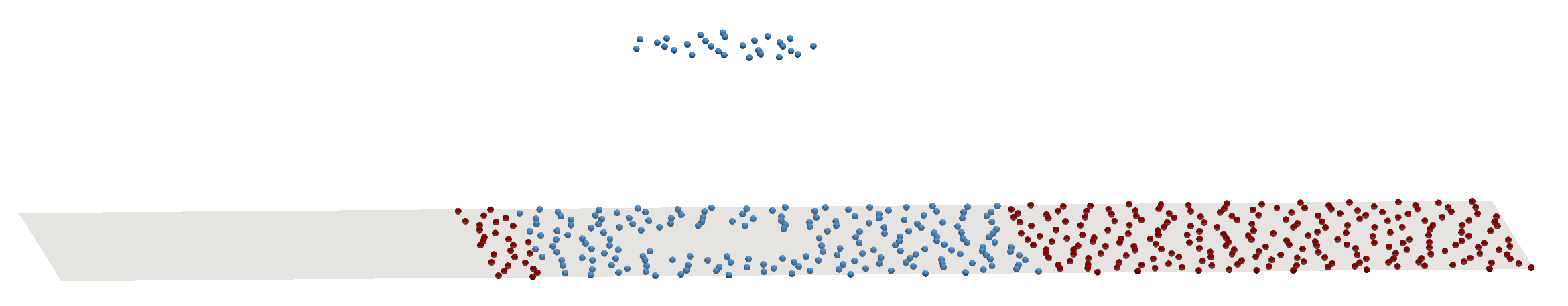}  \label{Fig:ARc}}\\
  \subfloat[$t=1.495$]{ \includegraphics[align=c,width=0.9\textwidth]{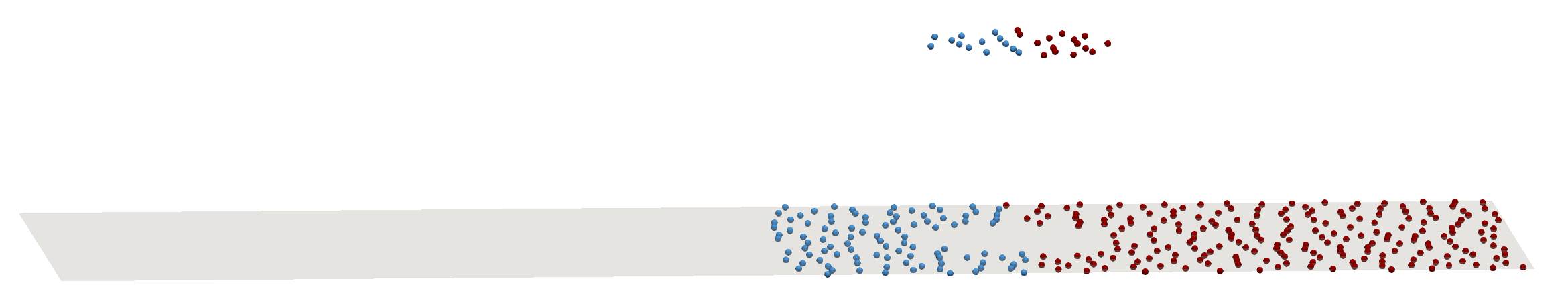}  \label{Fig:ARd}}\\
  \subfloat[$t=1.945$]{ \includegraphics[align=c,width=0.9\textwidth]{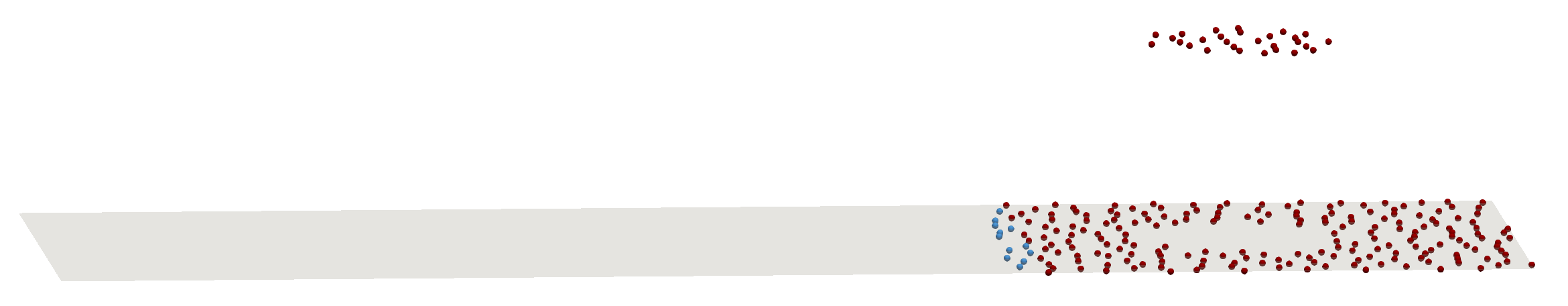}  \label{Fig:ARe}}
	\caption{Advection diffusion in uniform flow: Temperature profile for the pure advection case. 2D discretisation in maroon, 3D discretisation in blue. All particles with $T=0$ are shown on the surface, and all particles with $T>0$ are raised proportional to the value of $T$. For the 3D discretisation (blue), for the ease of visualizing the solution, only the boundary particles are shown with the temperature indicating the average temperature across the height of the discretization at that location. }
  \label{Fig:AdvectionResults}%
\end{figure}

The initial velocity $\vec{v} = (1,0,0)^T$ corresponds to fluid moving from $x=0$ to $x=3$, or left to right in Figure~\ref{Fig:AdvectionDomain}. Since there is no inflow of fluid, and a Lagrangian framework is used, no fluid is present on the left of the domain as the simulation progresses, see Figures~\ref{Fig:AdvectionDomain_b}--\ref{Fig:AdvectionDomain_d}. The initial temperature field is set to $0$ everywhere except a small circular region (see Figure~\ref{Fig:ARa})
\begin{equation}
\label{Eq:AdvDiff_IC}
    T(\vec{x},t=0) = 
    \left\{
    \begin{array}{ll}
        1, & \text{for } \| (x,y) - (x_0,y_0) \| \leq 0.2 \\
        0, & \text{elsewhere }\
    \end{array} 
    \right. ,
\end{equation}
where $(x,y)$ are the $x$ and $y$ coordinates of $\vec{x}$, and $x_0 = y_0 = 0.5$. Note that the temperature is constant in the depth direction $z$. Since the models produce a steady velocity equal to the initial velocity, the analytical solution to the temperature transport is given by Eq.\,\eqref{Eq:AdvDiff_IC} with $x_0 = 0.5 + t$ and $y_0 = 0.5$.

Since the advection is performed in a Lagrangian framework, no numerical diffusion is observed in either the 2D or the 3D model. The numerical results of the temperature profile are shown in Figure~\ref{Fig:AdvectionResults}. The figure illustrates that the transfer of data between the two models is exact to numerical precision. The final temperature at $t=2$ exactly matches the analytical solution. This forms a further verification of the data transfer between models during model change. A quantified comparison is presented in the next subsection. Note that in a more general scenario, when a discrete field varies along the depth of the domain, information will be lost while averaging from the 3D model to the 2D model. However, since the temperature field is constant along the depth of the domain in the present test case, there is no information loss due to model transition. 


%

\subsubsection*{Advection diffusion}~\\
In the pure advection test case considered above, since a Lagrangian framework is being used, there is no impact of derivative computation on the transport equation. I now extend this to a case with $\kappa \neq 0$, i.e. with diffusion of the temperature field. This will test the mechanism of the ghost particles as they work with derivative computation.

A similar setup to the pure advection case is considered, but with a larger domain. Fluid is flowing over the surface $[0,4] \times [0,3] \times [0]$ in the $xy$ plane. Model transition follows the same setup as the pure advection case. Initially, the 2D model is applied throughout the domain. Starting from the first time step, the 3D model is applied in the region $x \in [1,2]$. Due to the Lagrangian movement of the fluid, at every time step, 2D particles crossing $x=1$ are transitioned to the 3D model, and 3D particles crossing $x=2$ are transitioned to the 2D model. The same initial velocity of $\vec{v} = (1,0,0)^T$ is applied, with inviscid fluid without gravity, resulting in the same uniform and constant velocity as earlier. Since the numerical velocity remains $(1,0,0)^T$ in both the pure advection case above and the advection diffusion test case here, it shows that no spurious effects are introduced through the model transition or the data transfer mechanism using ghost particles. A diffusivity of $\kappa = 5 \times 10^{-3}$ is considered, with the initial temperature as specified in Eq.\,\eqref{Eq:AdvDiff_IC}.

\begin{figure}
  \centering
  \includegraphics[align=c,width=0.9\textwidth]{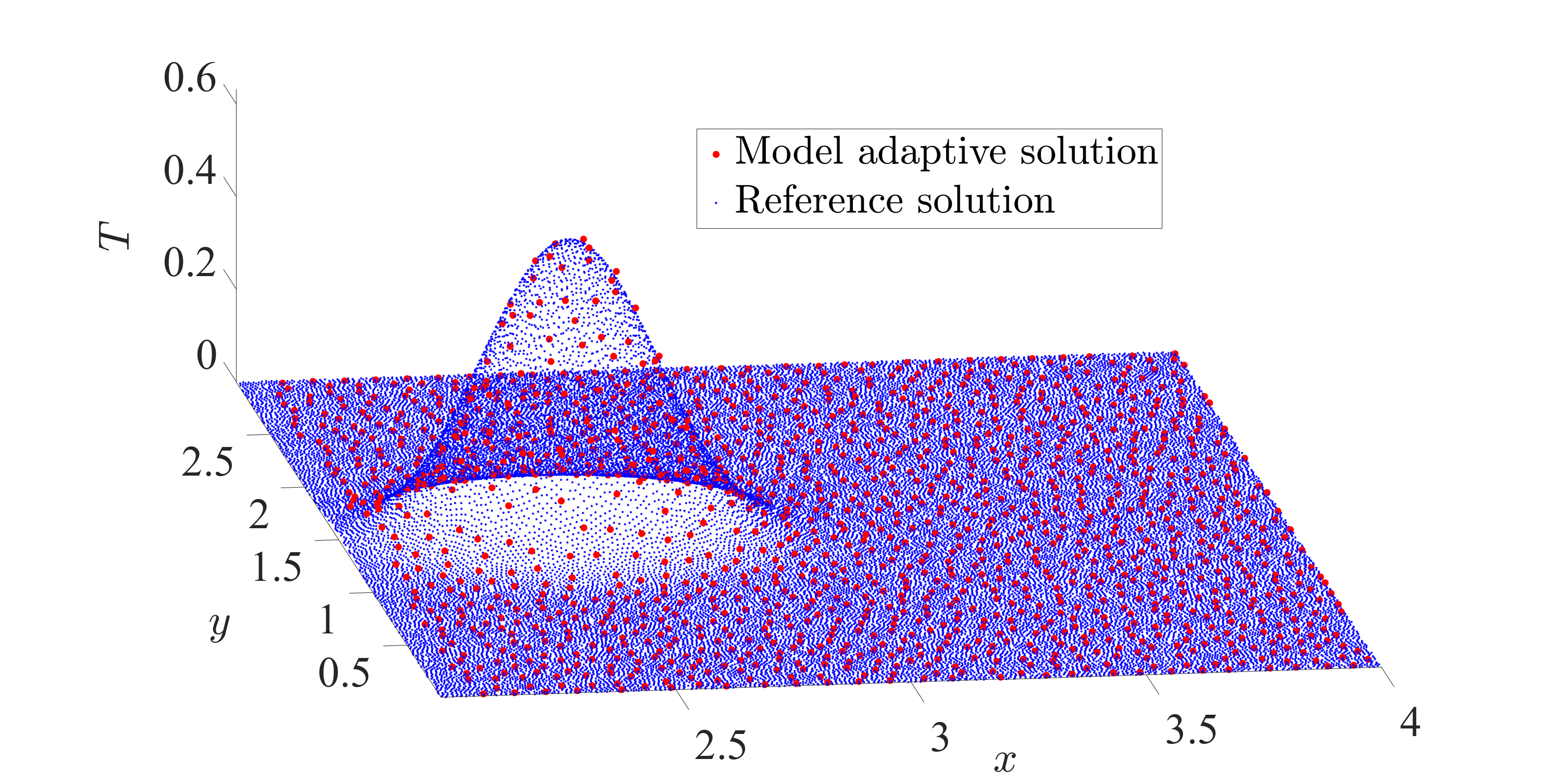}  
	\caption{Advection diffusion in uniform flow with $\kappa \neq 0$: Comparison of temperature profile at $t=2$ of the model adaptive method~(red) against a fine full 3D reference solution~(blue). The case of $h=0.2$ is shown for the model adaptive solution. See Figure~\ref{Fig:AdvDiff_Convergence} and Table~\ref{tab:AdvDiffErrors} for a quantified comparison with the reference solution.}
  \label{Fig:AdvDiff_Comparison}%
\end{figure}

Numerical results using the model adaptive framework are compared against a very fine 3D solution with $h = 0.04$ and $N = 644 860$ particles at initialization. For the model adaptive simulations, the same resolution is for both the 3D and 2D models. For $h=0.2$, resulting in $N^{3D} = 2234$ particles and $N^{2D} = 2199$ particles, the model adaptive solution is compared to fine 3D reference solution in Figure~\ref{Fig:AdvDiff_Comparison}. Note that the reference solution is constant along the depth ($z$) direction. Thus, only the boundary particles are shown in the figure to facilitate a visual comparison. The figure shows a good agreement between the reference and model adaptive solution, despite the use of significantly lesser particles in the model adaptive case. 

\begin{figure}
  \centering
  \includegraphics[align=c,width=0.9\textwidth]{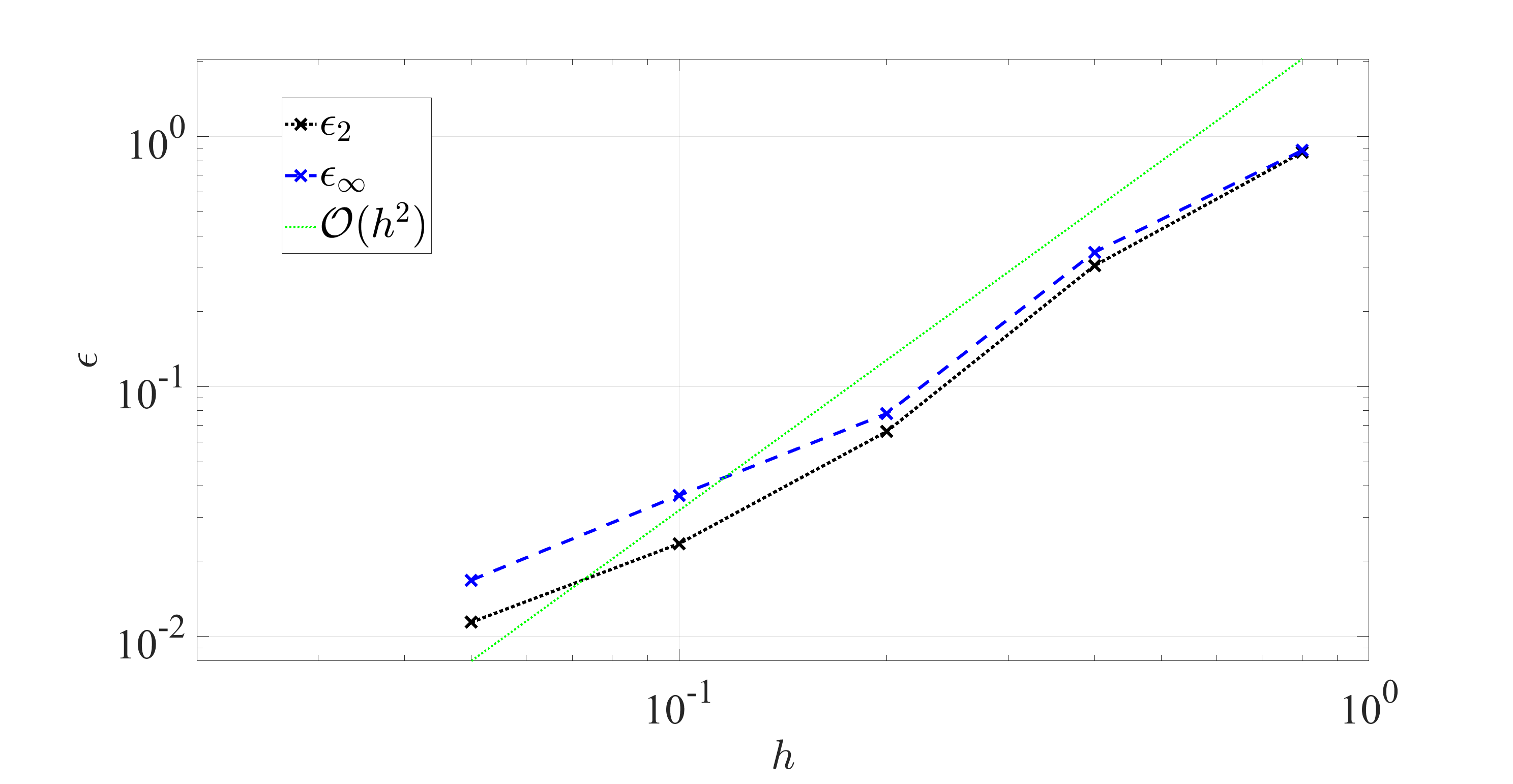}  
	\caption{Advection diffusion in uniform flow: Convergence of relative errors as the adaptive discretisation is refined for the $\kappa \neq 0$ case. Relative $L^2$ error $\epsilon_2$~(black) and relative $L^\infty$ error $\epsilon_\infty$~(blue) compared to a second order convergence rate~(green). The errors are tabulated in Table~\ref{tab:AdvDiffErrors}.}
  \label{Fig:AdvDiff_Convergence}%
\end{figure}

To quantify the difference in the model adaptive solution and the reference solution, I compute relative mean square and relative maximum errors as 
\begin{align}
    \epsilon_2 &= \frac{ \sum_{i=1}^N (T_i - T^{\text{ref}}_I )^2 }{ \sum_{i=1}^N (T^{\text{ref}}_I )^2 } \,, \label{Eq:ep2}\\ 
    \epsilon_{\infty} &= \frac{ \max_i(T_i - T^{\text{ref}}_I ) }{ \max_i(T^{\text{ref}}) }\,,\label{Eq:epInf}
\end{align}
where the summation $i$ is over all particles in the model adaptive solution at the end time of $t=2$, and $I$ is the closest particle in the reference solution to the particle $i$. Note that due to the unit velocity and Lagrangian framework, at the end time of $t=2$, the numerical domain consists only of 2D particles in the region $x \in [2,4]$, as shown in Figure~\ref{Fig:AdvDiff_Comparison}. A quantified error comparison and convergence study as the model adaptive domain is refined is plotted in Figure~\ref{Fig:AdvDiff_Convergence} and tabulated in Table~\ref{tab:AdvDiffErrors}. These results show a convergence of approximately second order, which matches the expected convergence rate due to the second order accurate GFDM discretization used for derivatives. This once again highlights that no spurious effects are introduced through the ghost particle or model transition mechanism.
\begin{table}
	\caption{Advection diffusion test case: Convergence of error in model adaptive solution when compared to a fine 3D reference solution. $h$ is the resolution used, $N^{3D}$ is the number of 3D particles, and $N^{2D}$ is the number of 2D particles, $N^{\text{total}} = N^{3D} + N^{2D}$ is the total number of particles. All particle numbers correspond to that at the end of the first time step, just after the 3D phase has been created. $\epsilon_2$ and  $\epsilon_{\infty}$ (see Eqs.\,\eqref{Eq:ep2},\eqref{Eq:epInf}) are the $L^2$ and $L^{\infty}$ errors respectively at $t=2$.}
	\centering
	\label{tab:AdvDiffErrors}
	{  
    \begin{tabular}{|c|r|r|r|c|c|}
	\hline
    $h$   & $N^{3D}$ &  $N^{2D}$  & $N^{\text{total}}$ & $\epsilon_2$  &  $\epsilon_{\infty}$  \\ 
	\hline \hline
$h = 0.8$   & $70$        &  $134$      & $204$  & $8.65 \times 10^{-1}$ &   $8.79 \times 10^{-1}$  \\\hline
$h = 0.4$   & $330$       &  $502$      & $832$  & $3.04 \times 10^{-1}$ &   $3.45 \times 10^{-1}$  \\\hline
$h = 0.2$   & $2\,234$    &  $2\,199$   & $4\,433$  & $6.61 \times 10^{-2}$ &   $7.79 \times 10^{-2}$  \\\hline
$h = 0.1$   & $14\,950$   &  $8\,837$   & $23\,787$  & $2.35 \times 10^{-2}$ &   $3.67 \times 10^{-2}$  \\\hline
$h = 0.05$  & $126\,533$  &  $37\,867$  & $164\,400$  & $1.14 \times 10^{-2}$ &   $1.68 \times 10^{-2}$  \\\hline%
	\end{tabular}}
\end{table}

For the 3D part of the domain, in the coarsest two resolutions considered, $h=0.4$ and $h=0.8$, the height is resolved with insufficient particles. In these cases, all 3D particles are either on the wall or the free surface, where boundary conditions would be applied. Thus, due to the lack of sufficient interior particles in these cases, there are insufficient particles to solve the PDEs. To resolve this, I solve the PDE and the boundary condition (BC) simultaneously on all particles in the 3D phase, $\text{PDE} + \zeta \text{BC} = 0$, for $\zeta = 0.3$. I refer to \cite{suchde2018meshfree} for more details on this. 

Note that the temperature profile is considered to be the same across the height of the domain in the 3D model. Thus, diffusion occurs in the same directions in the 2D and 3D models, only in and parallel to the $xy$ plane. In a more general case, variation within the height profile would occur in the 3D model, but would be averaged out in the 2D model. Such a difference is not considered in this test case. 

\subsection{Cleaning jet}
\label{sec:Cleaning}

\begin{figure}
  \centering
  \subfloat[Simulation setup]{
  \includegraphics[align=c,width=0.7\textwidth]{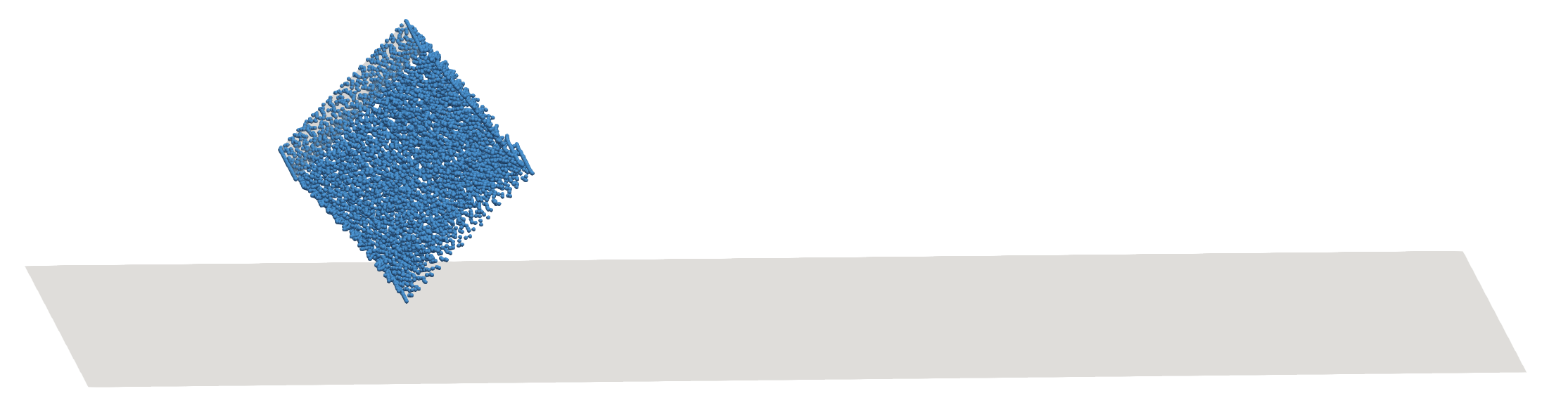} 
  \label{Fig:CleaningJet_Geometry_Main}
  }\\  
  \subfloat[Post-processing locations]{
  \includegraphics[align=c,width=0.7\textwidth]{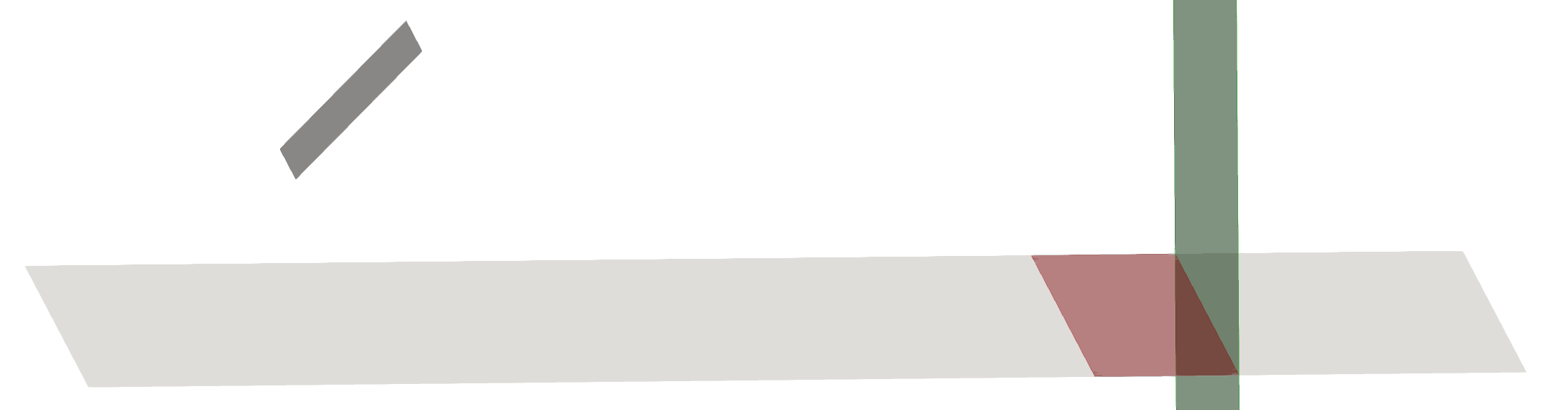}
  \label{Fig:CleaningJet_Geometry_PP}
  } 
  \caption{Cleaning jet: Illustration of test case and post-processing locations. Figure (a) shows the inflow (at the top left in each figure) of fluid (blue particles) which will hit the bottom surface being cleaned. Quantification of the results are done using virtual post-processing planes shown in Figure (b). The two quantities of interest studied are the flux of fluid passing through the vertical green plane, and the average normal stresses in the horizontal red region. }
  \label{Fig:CleaningJet_Geometry}%
\end{figure}

The previous two test cases considered model change in steady flow. I now test the full model adaptive scheme during unsteady flow. 
Consider a concept cleaning simulation. This test case is motivated by cleaning applications in the food industry, where large trays are cleaned in what is known as ``cleaning-in-place" processes. For the concept simulation considered here, I consider a flat surface being ``cleaned", without any contamination on it; and a jet of incoming fluid, as shown in Figure~\ref{Fig:CleaningJet_Geometry_Main}. As the jet of fluid hits the surface, it forms a thin layer of fluid. Modelling the thin layer of fluid is essential for cleaning applications, since it reduces the cleaning efficiency due to reduced surface stresses as compared to the hypothetical situation where no fluid film is present. 
Since the thickness of the film can be much smaller than the characteristic dimension of the jet inlet, the resolution needed for a purely 3D simulation is very fine. Since a thin film approximation is not valid at the inflow, both 2D and 3D models are required for this simulation. Furthermore, the regions where the 2D model will be needed can not be specified a-priori. Thus, this test case forms the perfect application for the model adaptive framework developed in the present work. 

\begin{figure}
  \centering
  %
  \includegraphics[align=c,width=0.7\textwidth]{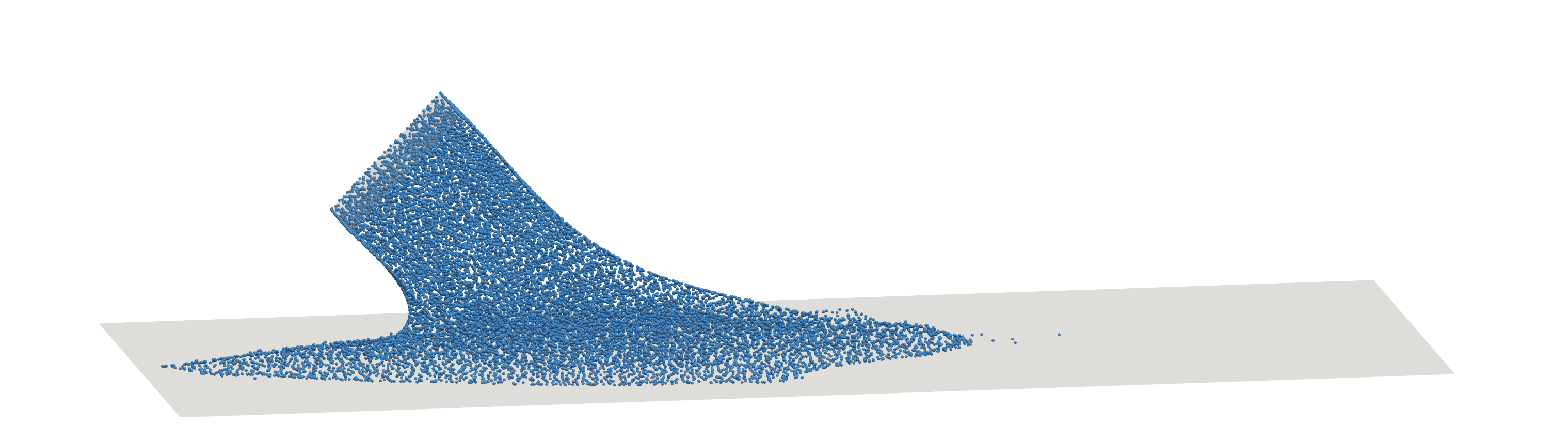}\\
  %
  \subfloat[$t=0.21$s. 3D only (top), and model adaptive (bottom). 3D model and dsicretization shown in blue, 2D in maroon.]
  {\includegraphics[align=c,width=0.7\textwidth]{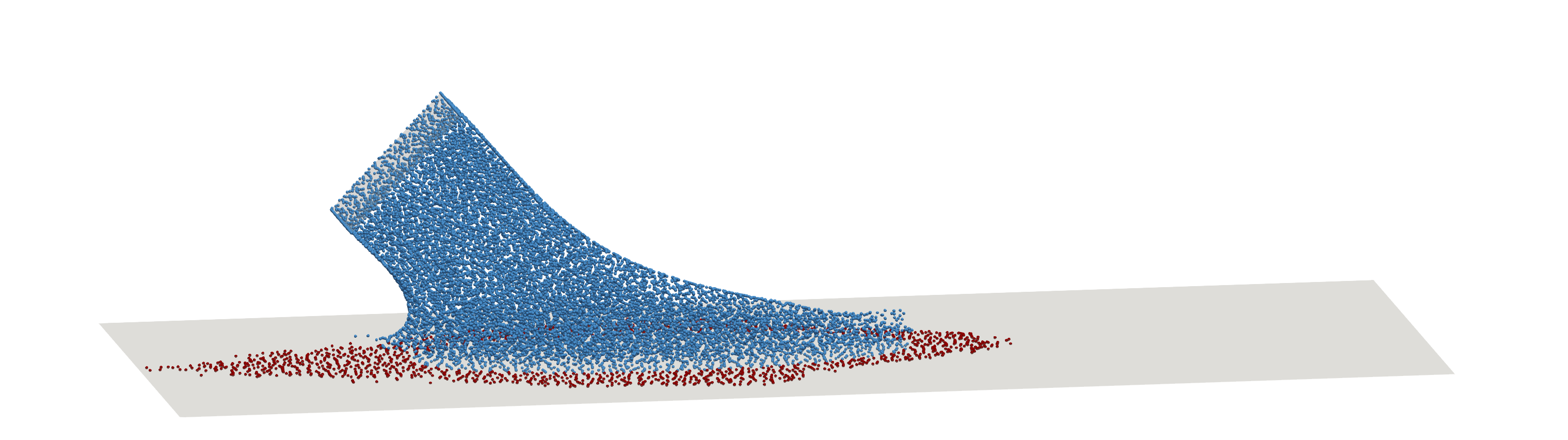}}\\
  %
  \includegraphics[align=c,width=0.7\textwidth]{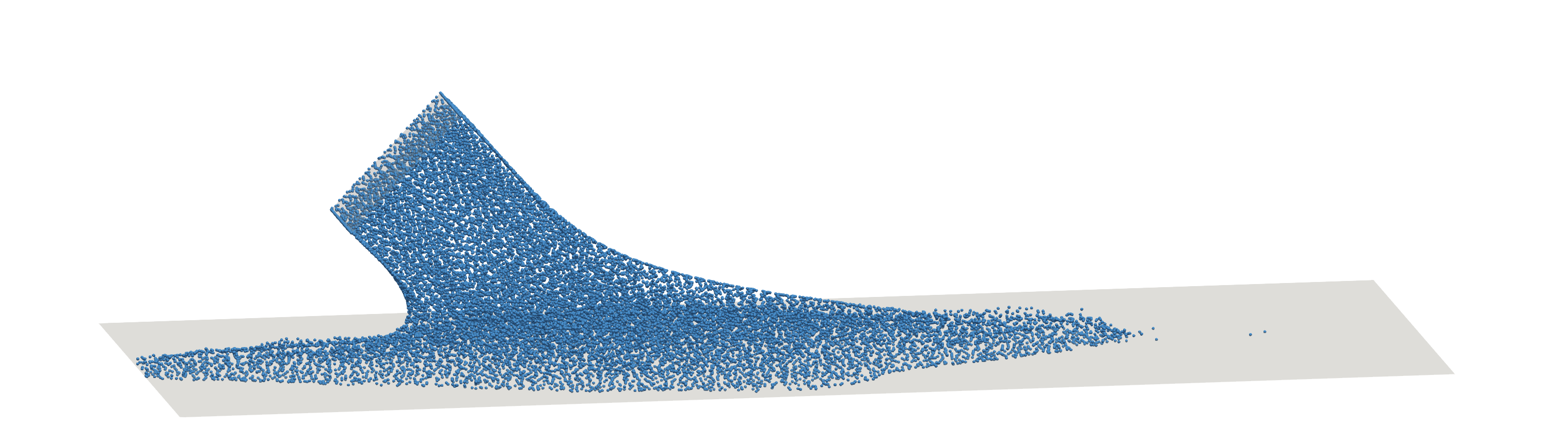}  \\ 
  %
  \subfloat[$t=0.26$s. 3D only (top), and model adaptive (bottom). 3D model and dsicretization shown in blue, 2D in maroon.]
  {\includegraphics[align=c,width=0.7\textwidth]{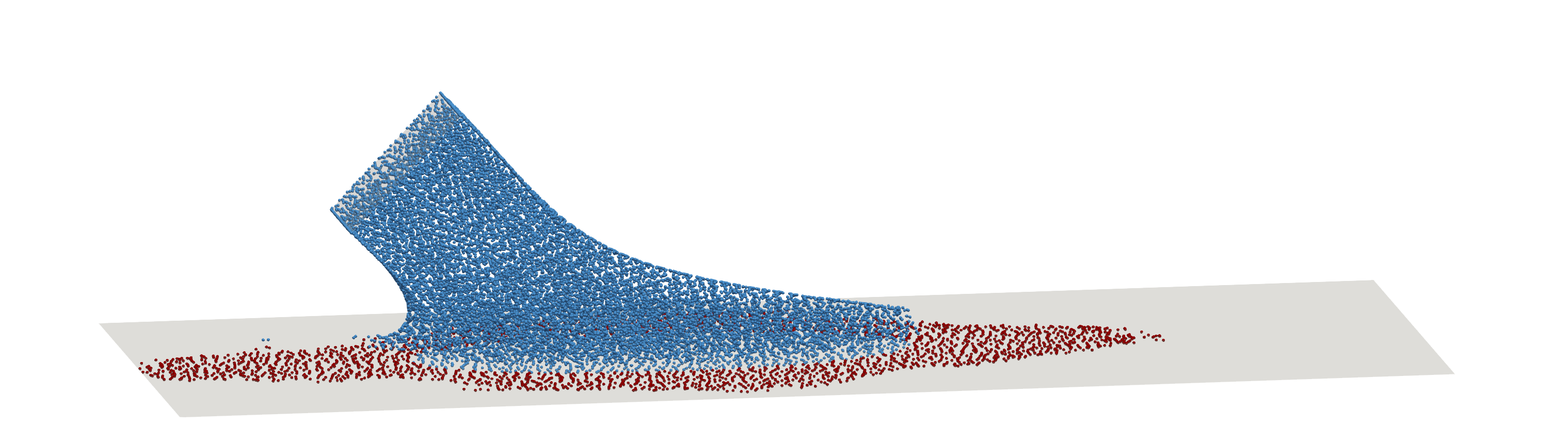}}\\        
  %
  \includegraphics[align=c,width=0.7\textwidth]{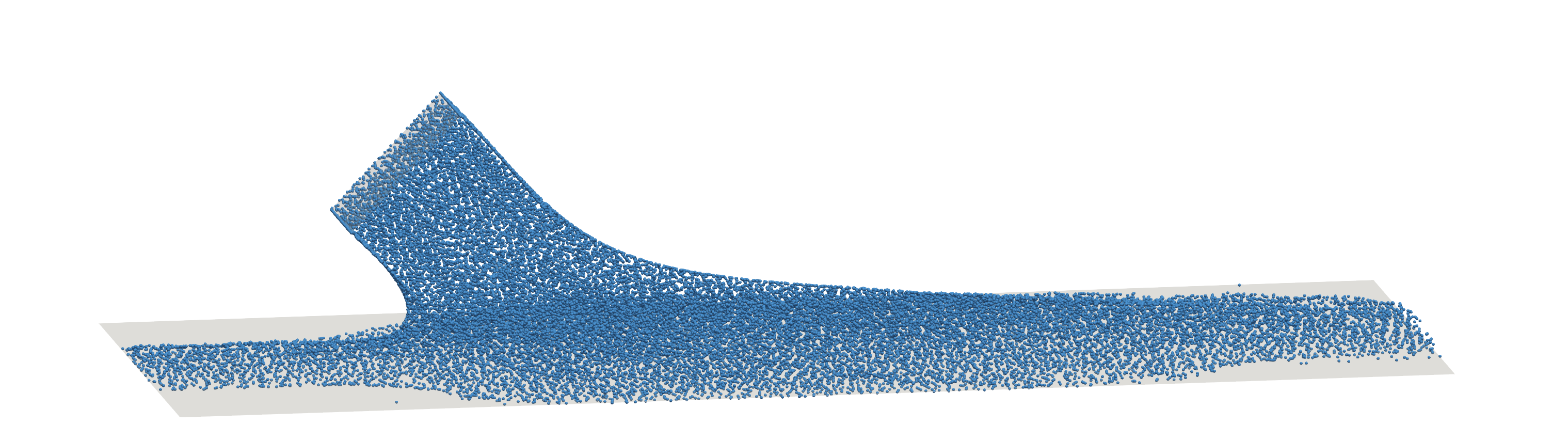}  \\
  %
  \subfloat[$t=0.4$s. 3D only (top), and model adaptive (bottom).3D model and dsicretization shown in blue, 2D in maroon.]
  {\includegraphics[align=c,width=0.7\textwidth]{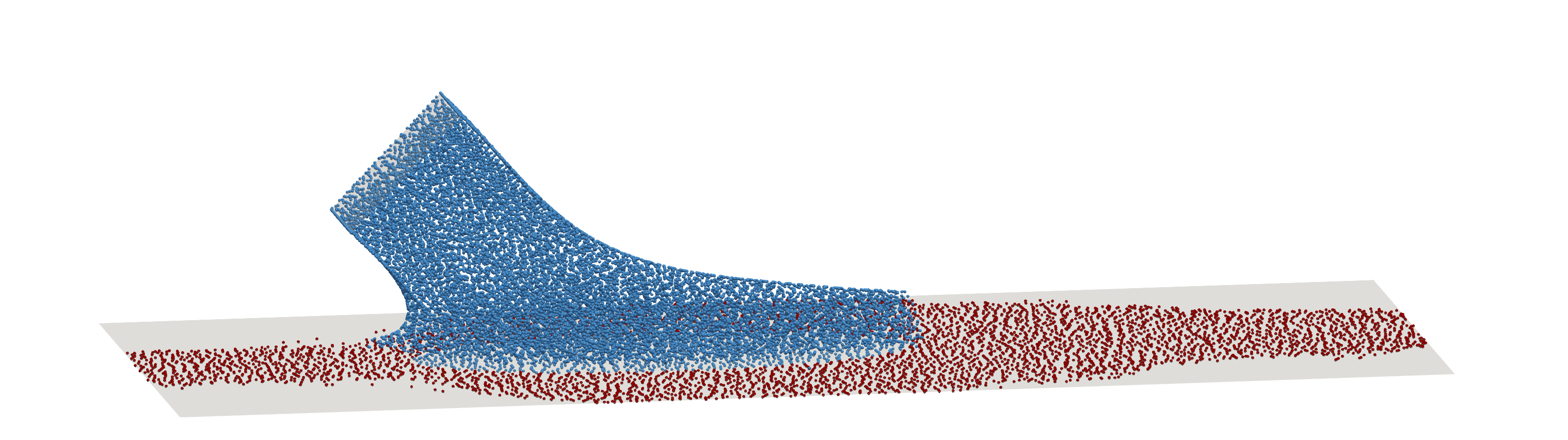}}
  \caption{Cleaning jet: Visual results comparisons between the 3D only simulations and the model adaptive (3D+2D) simulations. Particles in blue indicate the 3D phase, while particles in maroon indicate the 2D phase.}
  \label{Fig:CleaningJet_Results}%
\end{figure}

First, I have a look at the result qualitatively, to determine if the identification of thin films is done correctly. Figure~\ref{Fig:CleaningJet_Results} shows a comparison of a 3D only simulation with the model adaptive simulation at various time steps. For the model adaptive simulation, the same resolution is used for both the 3D and 2D phases. In the 3D only simulation, the resolution is refined near the surface to capture the flow in the thin fluid regions. For the model adaptive framework, the figure suggests a good behaviour of the PCA-based thin film detection method. At a slight distance from where the jet hits the surface, the presence of thin fluid films are detected, and the 3D model gets converted to the 2D model. 

\begin{figure}
  \centering
  \subfloat[Instantaneous flux $q$, see Eq.\,\eqref{Eq:InstFlux}.]
  {\includegraphics[align=c,width=0.9\textwidth]{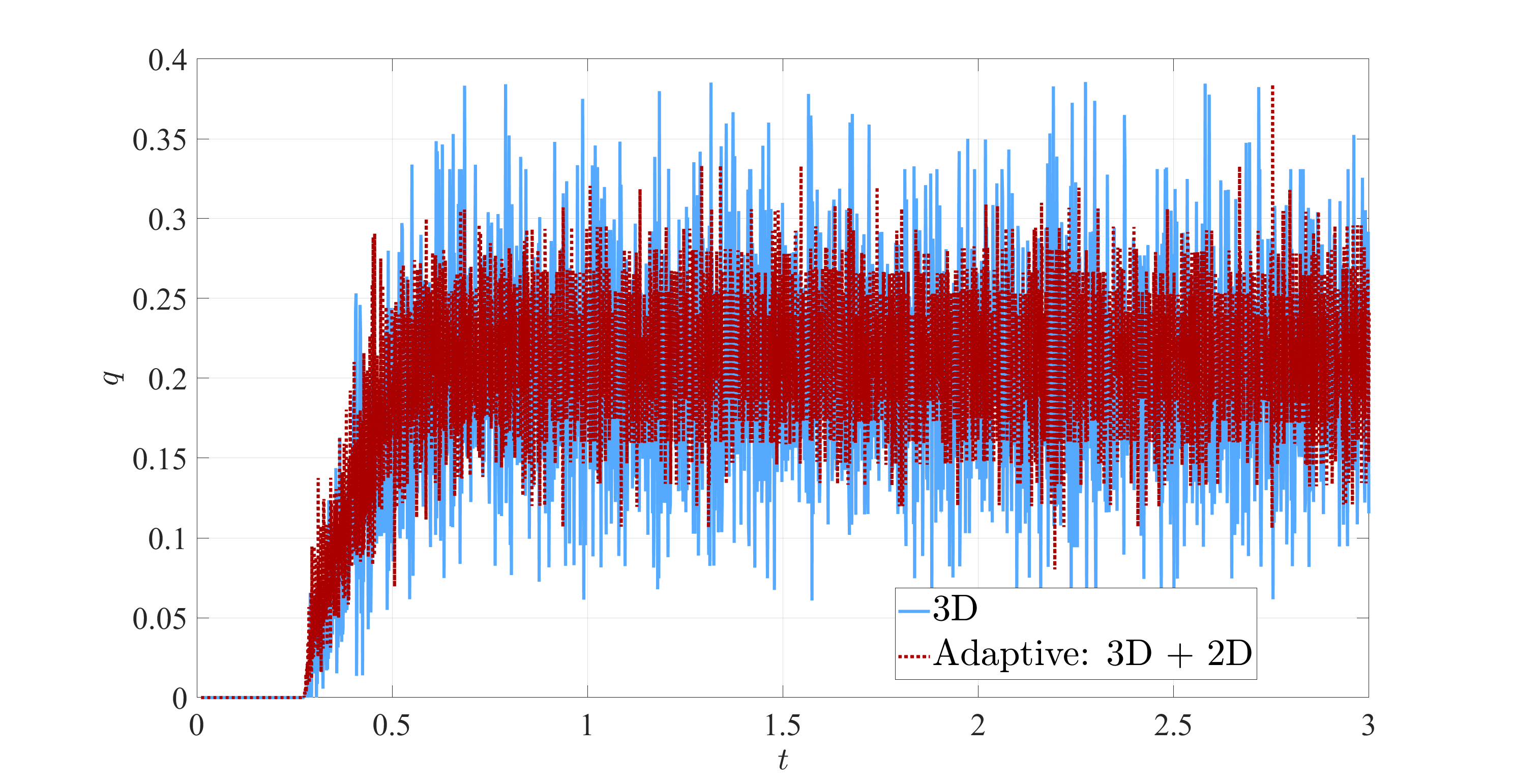}}\\
  \subfloat[Cumulative flux $Q$, see Eq.\,\eqref{Eq:TotFlux}.]
  {\includegraphics[align=c,width=0.9\textwidth]{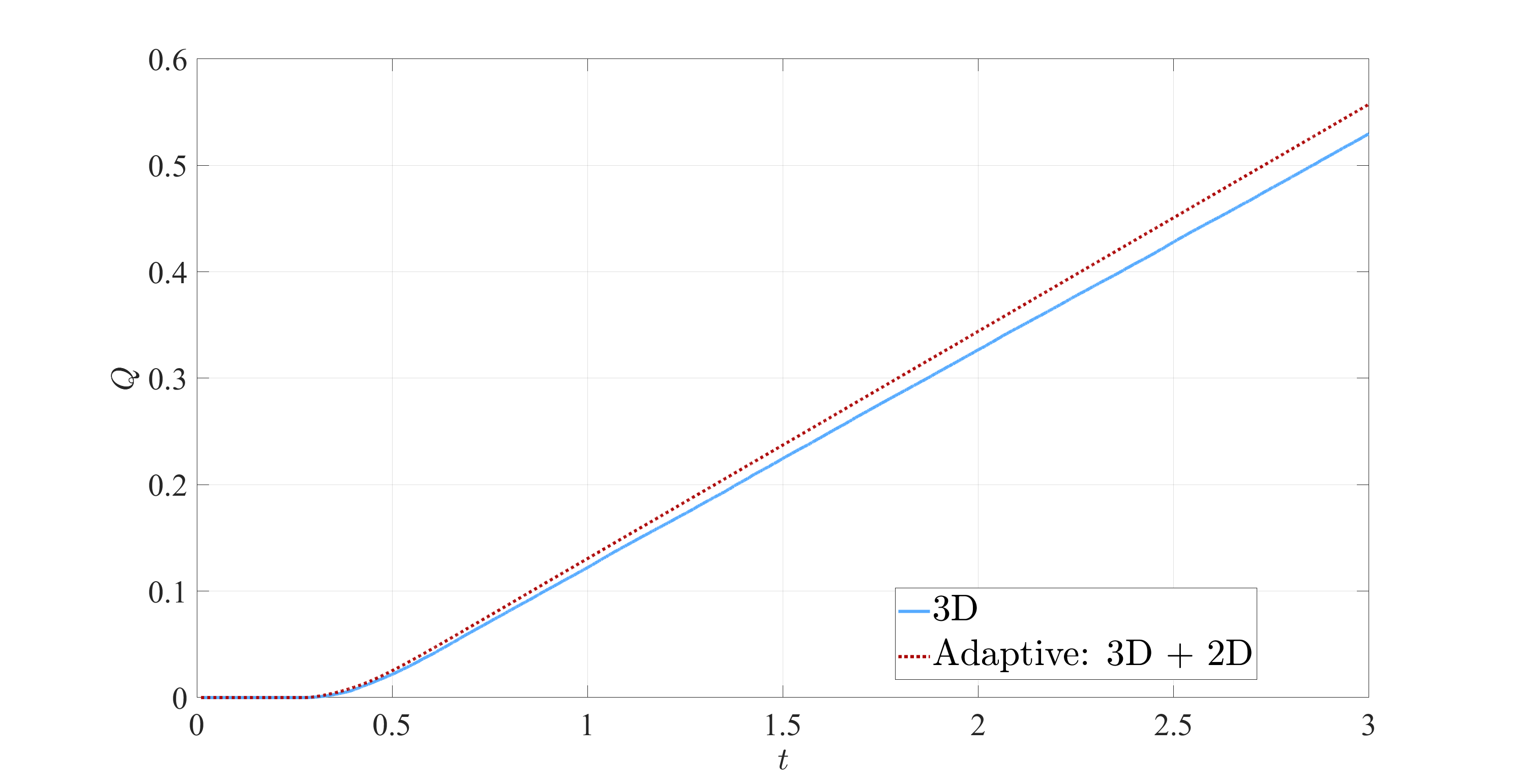}}
	\caption{Cleaning jet: Comparison of fluid flux through a virtual plane for the fine 3D and model adaptive (3D+2D) simulations. See Figure~\ref{Fig:CleaningJet_Geometry_PP} for the location of the virtual post-processing plane.}
  \label{Fig:Cleaning_Flux}%
\end{figure}

To quantify the difference between the 3D only simulation and the hybrid model adaptive results, I measure the fluid flux across a fictitious vertical plane shown in green in Figure~\ref{Fig:CleaningJet_Geometry_PP}. The instantaneous flux $q$ through the virtual plane at a time $t$, and the time cumulative flux $Q$ indicating the total fluid flowing through the plane until time $t$ are measured as
\begin{align}
    q &= \int_{\partial \Omega_{\text{virtual}}} \Vec{v} \cdot \vec{n} \,dA \,,\label{Eq:InstFlux}\\
    Q &= \int_0^t q(\tau) \, d\tau \,.\label{Eq:TotFlux}
\end{align}
Numerically, $q$ is computed as a sum over all particles crossing the virtual plane. A comparison of results for the 3D and model adaptive simulations, for both the instantaneous and cumulative flux, are shown in Figure~\ref{Fig:Cleaning_Flux}. The figure suggests a good match between the model adaptive results and the fine 3D results. For the instantaneous flux, both simulations exhibit a fluctuation around the same mean value of approximately $0.2$. The model adaptive results show lesser fluctuations for instantaneous flux than the pure 3D model. This is expected due to the depth averaged flow profile in the 2D model. 
As the error accumulates, the maximum relative error in $Q$ at the end of the simulation time is only $4.2 \%$.
%

\begin{figure}
  \centering
  \includegraphics[align=c,width=0.9\textwidth]{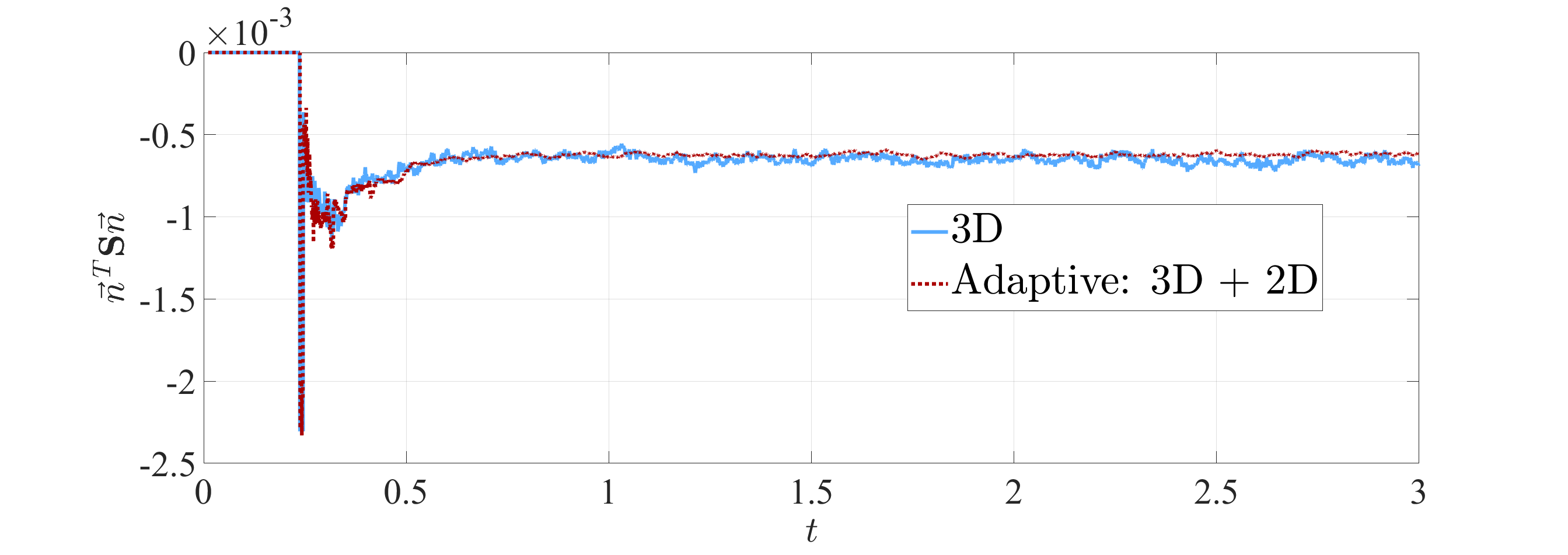}
	\caption{Cleaning jet: Comparison of cleaning effect of the 3D and model adaptive (3D+2D) simulations, given by the scalar component of the viscous stress perpendicular to the surface. }
  \label{Fig:Cleaning_Stress}%
\end{figure}

Since this test case is motivated by a cleaning application, for a further quantified comparison between the 3D and model adaptive simulations, I propose an indicator of how well the fluid will clean the surface. I compare the scalar component of the viscous stress perpendicular to the surface, defined as $\vec{n}^{T} \mathbf{S} \vec{n}$ for viscous stress $\mathbf{S} = \eta \left( \nabla\vec{v} + (\nabla\vec{v})^T \right)$. This represents a notion of comparison of the cleaning effect of the jet on the surface. Note that this also forms the normal component of the traction. Rather than considering the normal traction at a single location, I consider an average over a small region of the surface, as shown in red in Figure~\ref{Fig:CleaningJet_Geometry_PP}. The average value of $\vec{n}^{T} \mathbf{S} \vec{n}$ in this region is shown in Figure~\ref{Fig:Cleaning_Stress} for both 3D and model adaptive simulations. 
Once again, both results are very similar. The initial value of $0$ indicates that fluid has not yet reached this post-processing zone. The highest deviation between the two approaches is observed when the fluid first reaches this zone, where the maximum deviation is $8.1 \%$ and there are high fluctuations in the normal traction.
%

Since the focus of the present work is on the model adaptivity, several important fluid effects for these applications have not been considered. Most importantly, surface tension and wetting angle effects have been neglected. 

\subsection{Application: Automotive water crossing}
\label{sec:WaterCrossing}

\begin{figure}
  \centering
  %
  \includegraphics[align=c,width=0.92\textwidth]{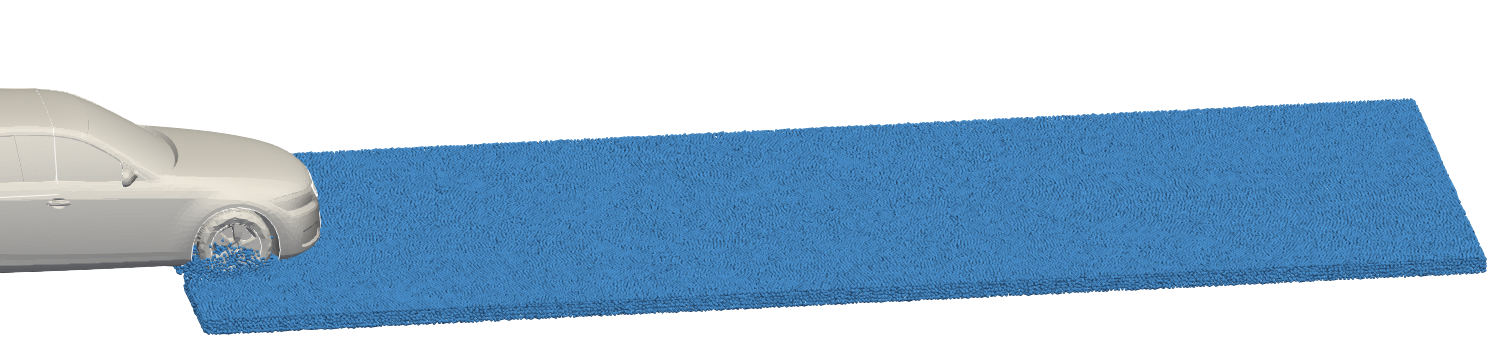} \\
  %
  \subfloat[$t=0.07$s. 3D only (top), and model adaptive (bottom). 3D model and discretization shown in blue, 2D in maroon.]
  {\includegraphics[align=c,width=0.92\textwidth]{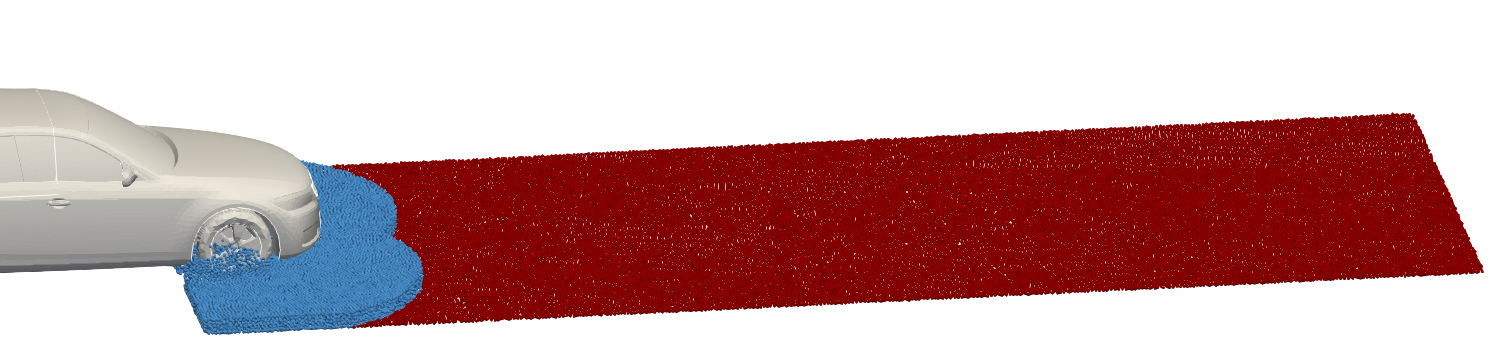}}\\
  %
  \includegraphics[align=c,width=0.92\textwidth]{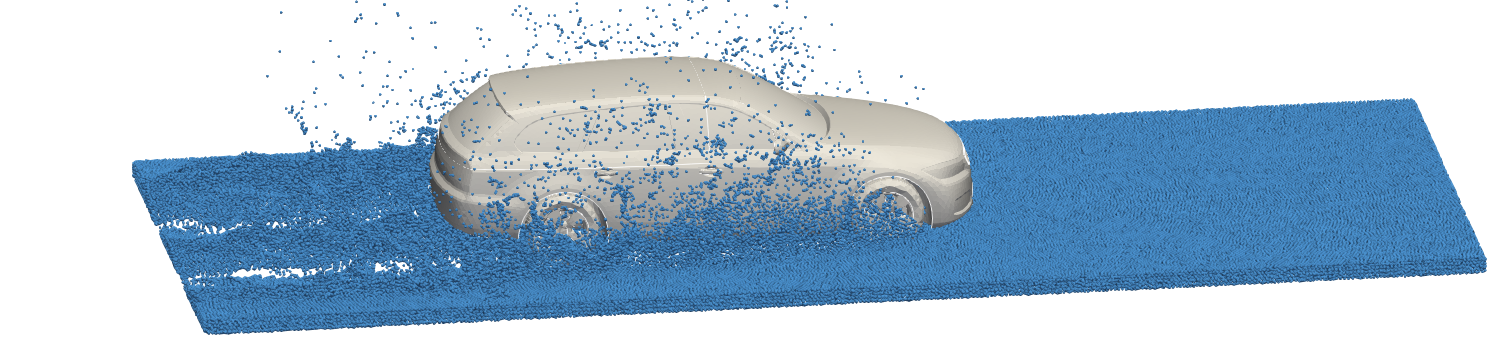}  \\ 
  %
  \subfloat[$t=0.57$s. 3D only (top), and model adaptive (bottom). 3D model and discretization shown in blue, 2D in maroon.]
  {\includegraphics[align=c,width=0.92\textwidth]{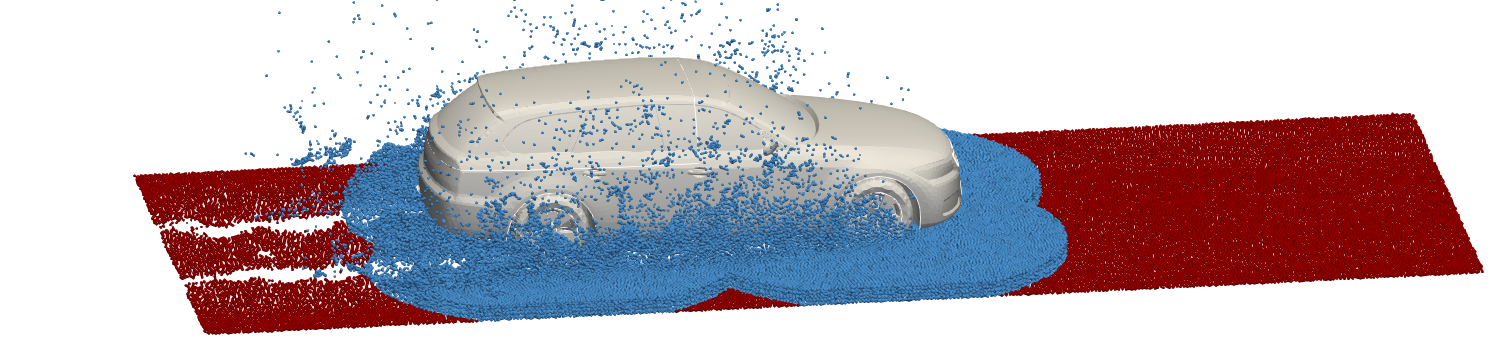}}\\        
  %
  \includegraphics[align=c,width=0.92\textwidth]{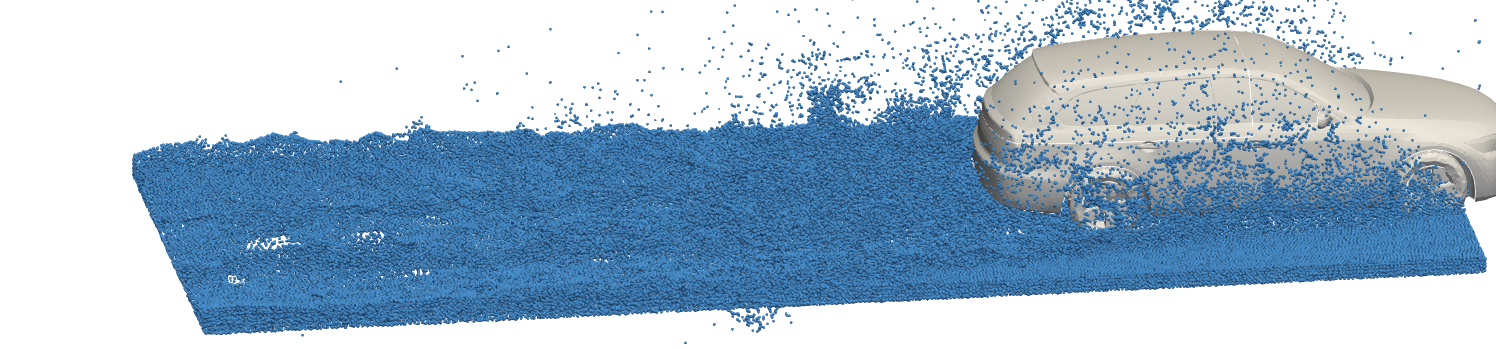}  \\
  %
  \subfloat[$t=1$s. 3D only (top), and model adaptive (bottom). 3D model and discretization shown in blue, 2D in maroon.]
  {\includegraphics[align=c,width=0.92\textwidth]{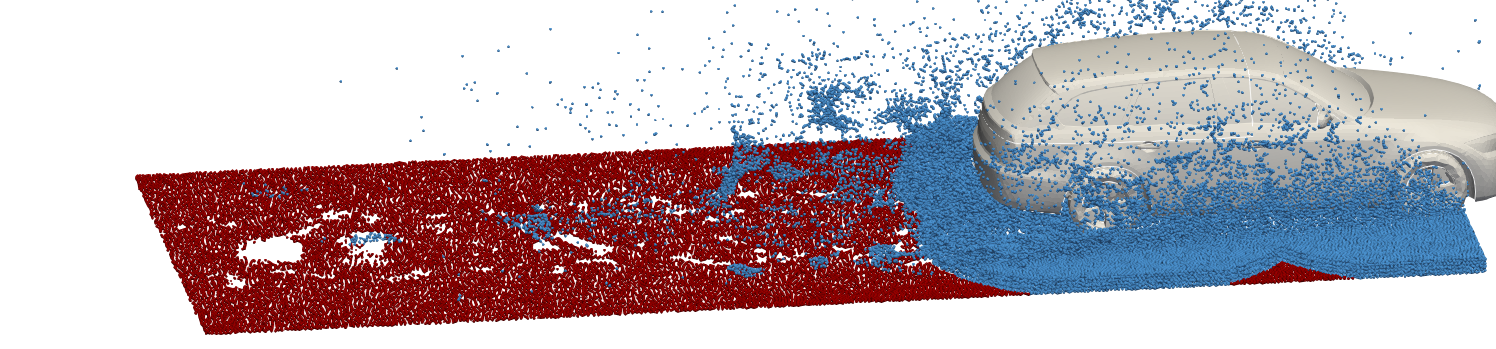}}
	\caption{Automotive water crossing: Illustration of user-defined transition criteria for transitioning between the 2D thin film and 3D bulk models. Here, I use a minimum distance from the four car wheels as the criterion for model change. particles in blue indicate the 3D phase, while particles in maroon indicate the 2D phase.}
  \label{Fig:WaterCrossing}%
\end{figure}

I now present an application of the model adaptive framework introduced in the present work on an industrial test case. Consider a car crossing a shallow pool of water of constant depth, see \cite{suchde2023volume, zhang20233d}. The study of water wading depth, or the depth of water through which a car can safely cross, has become very important in vehicular design. As shown in Figure~\ref{Fig:WaterCrossing}, the simulation domain consists of a long shallow pool of water, about $3$ times the length of the car. The crucial part of the simulation domain is the fluid just around the car. Fluid in the far field plays a negligible role on the main quantities of interest: the traction on the wheels, and the water spray patterns. For more details on this application, and for mass conservation considerations, see \cite{suchde2023volume}.

For such a scenario, adaptive refinement significantly helps in speeding up simulations. However, a significant drawback in meshfree methods is the lack of anisotropic resolutions. This introduces a minimum resolution needed to resolve the thickness of the film, which is needed throughout the domain. 
To overcome this drawback in using adaptive refinement, I propose the use of adaptive model selection. The automatic detection of model adaptivity presented in Section~\ref{sec:Where} is supplemented with a user-defined criterion based on the distance of the fluid to the car. Particles sufficiently close to the car are solved with the 3D model, while those far away are solved with the 2D model. As a result, when the car moves across the channel (from left to right in Figure~\ref{Fig:WaterCrossing}), at every time step, 2D particles coming near the car are transitioned to the 3D model, and 3D particles far away from the car are shifted to the 2D model. 

The height of the pool is $12 \text{cm}$, and the car crosses the pool with a uniform speed of $40 \text{km/h}$. The car is assumed to be a rigid body, except the wheels and tyres, each of which rotate about their centre point with an angular velocity that matches the prescribed liner velocity of the car. Snapshots of the numerical results of the 3D simulation and model adaptive simulation are shown in Figure~\ref{Fig:WaterCrossing}. The figure shows that both methods produce very similar results. The fluid is at rest in front of the car. Thus, no information is lost by applying the 2D model ahead of the car (Figure~\ref{Fig:WaterCrossing} shows where the 2D model is being used). As a result, the 3D region in the model adaptive simulation perfectly matches the corresponding region in the 3D only simulation. All quantities of interest measured around the car are exactly the same in both simulations. This was verified by checking the evolution of the pressure of the water on the wheels, and the spray of water hitting the car, for three resolutions each. 


The results of the two approaches (3D and model adaptive) only differ in the wake of the car. It is important to note here that typical quantities of interest in water crossing applications, like the traction on the wheels and the fluid spray on the car, are measured only around the car, with no influence of the flow patterns in the wake.

The number of particles in the discrete domain and the simulation time for simulating $1$ second of physical time is tabulated in Table~\ref{tab:WaterCrossing} for three different resolutions each of the 3D only approach and the model adaptive approach. In the model adaptive simulation, the same resolution is used for the both the 3D and 2D phases. Table~\ref{tab:WaterCrossing} shows that the coarsest simulation considered results in a speed-up of a factor of $1.7$ for the model adaptive approach. As the number of particles increases, the gap in the simulation time between the approaches increases. For the finest simulation of $h=0.1$, the model adaptive framework introduced in this work results in a speed-up of a factor of $3.2$, while maintaining the same flow results around the car.
\begin{table}
	\caption{Automotive water crossing: Number of points $N$, and computational time required $t_{\text{comp}}$ (in minutes) for a simulated time of $1$ second. All simulation are run in parallel using $4$ MPI processes. The number of points is reported at $t=0.5$.}
	\centering
	\label{tab:WaterCrossing}
	{  
    \begin{tabular}{|c||r|r||r|r|r|r|}
	\hline
        & \multicolumn{2}{c||}{3D only}    & \multicolumn{4}{c|}{Model adaptive (3D+2D)}  \\ 
    $h$   & $N^{3D}$ &  $t_{\text{comp}}$  & $N^{3D}$ &  $N^{2D}$  & $N^{\text{total}}$ & $t_{\text{comp}}$  \\ 
	\hline
$h = 0.2$   & $30\,497$   &  $12$   & $15\,956$  & $6\,453$ &    $22\,409$  & $7$ \\\hline
$h = 0.15$  & $63\,468$   &  $33$   & $30\,488$  & $11\,086$ &   $41\,574$  & $16$ \\\hline
$h = 0.1$   & $195\,201$  &  $181$  & $81\,202$  & $23\,351$ &   $104\,553$ & $56$ \\\hline%
	\end{tabular}}
\end{table}
%


\section{Limitations and open questions}
\label{sec:Limitations}

While the results shown in Section~\ref{sec:Results} suggest that the model adaptive framework presented here is very promising, there are several limitations and open questions that must be discussed. 

\subsubsection*{Model differences}
\begin{itemize}
    \item The differences in the assumptions of the two models are crucial in understanding the capabilities of the presented model adaptive framework. For instance, the 2D thin film model is derived under the assumption of a quadratic velocity profile \cite{bharadwaj2022discrete}. However, while making the 2D to 3D transition, the presented framework enforces a uniform velocity profile (with slip boundary conditions), rather than a quadratic profile. 
    \item In a general scenario, information is lost when depth averaging from the 3D model to the 2D model. This raises the question as to when does this transition result in an error accumulation that is unacceptable. Further tests need to be done to properly test the limits of this work.  
\end{itemize}

\subsubsection*{Detecting model transition}
\begin{itemize}
    \item For the identification of thin layers, a further study is needed to determine the best approach. The local PCA presented in Section~\ref{sec:Where} forms a sufficient first model, but requires further testing to determine its limits and optimal parameter values for $\epsilon_\lambda^{\text{max}}$ and $\theta^*$. We also need quantifiable tests to compare different methods of thin film detection. Further testing of the PCA approach is needed before a claim can be made concerning the robustness of the automatic model transition detection. Weighted PCA approaches \cite{Castillo2013}, and related work on the rupture of thin films \cite{chirco2022manifold} should be investigated as alternatives to the PCA approach presented here. 
    



    \item In the context of h-adaptivity, the use of error indicators and estimators to decide when to refine or coarsen a discretization has been shown to be a very robust and effective approach (see, for example, \cite{jacquemin2023smart, zienkiewicz2006background}). This raises the question if an error-based approach to change models during model adaptivity could be devised to either replace or compliment the heuristic based approach proposed here.
\end{itemize}

\subsubsection*{Data communication}~\\
While the buffer zone approach using ghost particles introduced in Section~\ref{sec:DataComm} works very well, as shown in Section~\ref{sec:Results}, a significant issue of this is that is hampers code maintainability. The addition of ghost particles handling to every instance of derivative computation makes the code cumbersome to maintain, especially when there are multiple developers. I thus wish to investigate alternative approaches to data communication without the use of ghost particles. 


 


\section{Conclusion}
\label{sec:Conclusion}

I presented a novel adaptive procedure to couple a 3D Navier--Stokes model with a pseudo-2D thin film flow model. By analysing the principal components of a local neighbourhood, the proposed algorithm detected when and where a thin film model would be applicable. Lagrangian particles were used as the domain discretization in both models. When the need for model change from 3D to 2D or vice versa is detected for a particle, all required data is transferred between models, and the required change in discretization (3D vs 2D) is also performed automatically. When particles governed by different models lie beside each other, data is communicated using a two-way coupling between them using a buffer region where both models are solved. The buffer region is populated with ghost particles that assist in data communication. Due to the dynamic nature of the typical applications of the model adaptive framework, ghost particles need to be created at every time step. Numerical results show a good comparison of the adaptive framework with a pure 3D bulk flow situation, with a significant time speed up. My future work in this direction will be towards investigating the limitations listed in Section~\ref{sec:Limitations}, and extending the work to flow on curved and moving surfaces.

While this work only considered the adaptive transition from a 3D flow model to a 2D one. This lays the foundation for an adaptive selection between any two models of the same phenomenon, in a vast variety of scenarios. This forms a crucial part of my ongoing work.  


\section*{Acknowledgements}

Pratik Suchde would like to acknowledge partial support from  the  European  Union's  Horizon  2020 research  and  innovation programme under the Marie Skłodowska-Curie Actions grant agreement No. 892761 ``SURFING". 
Pratik Suchde would also like to acknowledge funding from the Institute of Advanced Studies, University of Luxembourg, under the AUDACITY programme grant ``ADONIS".


\bibliographystyle{abbrv}
\bibliography{./ReferencesOthers, %
./ReferencesMeshfree}
\end{document}